\documentclass[aps,pre,onecolumn,tightenlines,final,letterpaper,notitlepage,superscriptaddress,longbibliography]{revtex4-1}

\usepackage[utf8]{inputenc}
\usepackage{graphicx}
\usepackage{amsmath,amssymb,amsthm}
\usepackage{txfonts}
\usepackage[inline]{enumitem}
\usepackage{hyperref}
\graphicspath{{./figs/}}

\begin{document}

\author{Laurent H\'{e}bert-Dufresne}
\affiliation{Vermont Complex Systems Center, University of Vermont, Burlington, VT 05405 }
\affiliation{Department of Computer Science, University of Vermont, Burlington, VT 05405 }
\affiliation{D\'epartement de physique, de g\'enie physique et d'optique,
Universit\'e Laval, Qu\'ebec (Qu\'ebec), Canada G1V 0A6}
\author{Benjamin M. Althouse}
\affiliation{Institute for Disease Modeling, Bellevue, WA}
\affiliation{University of Washington, Seattle, WA}
\affiliation{New Mexico State University, Las Cruces, NM}
\author{Samuel V. Scarpino}
\affiliation{Network Science Institute, Northeastern University, Boston, MA, 02115}
\affiliation{Department of Marine \& Environmental Sciences, Northeastern University, Boston, MA, 02115}
\affiliation{Department of Physics, Northeastern University, Boston, MA, 02115}
\affiliation{Department of Health Sciences, Northeastern University, Boston, MA, 02115}
\affiliation{ISI Foundation, Turin, 10126, Italy}
\author{Antoine Allard}
\affiliation{D\'epartement de physique, de g\'enie physique et d'optique,
Universit\'e Laval, Qu\'ebec (Qu\'ebec), Canada G1V 0A6}
\affiliation{Centre interdisciplinaire en mod\'elisation math\'ematique, Universit\'e Laval, Qu\'ebec (Qu\'ebec), Canada G1V 0A6}

\title{Beyond $R_0$: Heterogeneity in secondary infections and probabilistic epidemic forecasting}

\begin{abstract}
  The basic reproductive number --- $R_0$ --- is one of the most common and most commonly misapplied numbers in public health.  
  %Nevertheless, estimating $R_0$ for every transmissible pathogen, emerging or endemic, remains a priority for epidemiologists. %~\cite{Diekmann1995}.
  Although often used to compare outbreaks and forecast pandemic risk, this single number belies the complexity that two different pathogens can exhibit, even when they have the same $R_0$~\cite{Lloyd-Smith2005, Bansal2007, Vizi2017}.
  Here, we show how to predict outbreak size using estimates of the distribution of secondary infections, leveraging both its average $R_0$ and the underlying heterogeneity.  
  To do so, we reformulate and extend a classic result from random network theory~\cite{Newman2001} that relies on contact tracing data to simultaneously determine the first moment ($R_0$) and the higher moments (representing the heterogeneity) in the distribution of secondary infections.  
  Further, we show the different ways in which this framework can be implemented in the data-scarce reality of emerging pathogens.  
  Lastly, we demonstrate that without data on the heterogeneity in secondary infections for emerging infectious diseases like COVID-19, the uncertainty in outbreak size ranges dramatically. %, in the case of COVID-19 from 10-70\% of susceptible individuals. 
  Taken together, our work highlights the critical need for contact tracing during emerging infectious disease outbreaks and the need to look beyond $R_0$ when predicting epidemic size.
\end{abstract}

\maketitle

\section{Introduction}
In 1918, a typical individual infected with influenza transmitted the virus to between one and two of their social contacts~\cite{Biggerstaff2014}, giving a value of the basic reproductive number -- $R_0$, the number of secondary infections in a completely susceptible population -- between one and two.  These are similar to values of $R_0$ for the 2014 West Africa Ebola virus outbreak, but most individuals infected with Ebola virus gave rise to zero additional infections, while a few gave rise to more than 10~\cite{WHOEbolaResponseTeam2014, Althaus2014}.  Moreover, Ebola virus disease infected a tenth of one percent of the number of individuals believed to have been infected by the 1918 Influenza virus~\cite{Mills2004,Kaner2016}.  While improvements in healthcare and public health measures, as well as changes in human behavior, partially explain the massive discrepancy between Ebola virus disease in 2014 and influenza in 1918~\cite{Bootsma2007}, there is another critical difference between these two diseases: heterogeneity in the number secondary cases resulting from a single infected individual.  Here, we demonstrate analytically that quantifying the variability in the number of secondary infections is critically important for quantifying the transmission risk of novel pathogens.% and further show how a lack of publicly available contact tracing data on cases of novel coronavirus (COVID-19) prevents the global public health community from determining the true pandemic risk of this novel virus.

The basic reproduction number of an epidemic, $R_0$, is the expected number of secondary cases (note, we use the word ``case'' in a generic sense to represent any infection, even if too mild to meet the clinical case definition) produced by a primary case over the course of their infectious period in a completely susceptible population~\cite{Diekmann1995}.  It is a simple metric that is commonly used to describe and compare the transmissibilty of emerging and endemic pathogens~\cite{nyt}. If $R_0 = 2$, one case turns to two, on average, and two turn to four as the epidemic grows.  Conversely, the epidemic will die out if $R_0 < 1$.

Almost 100 years ago, work from Kermack and McKendrick~\cite{Kermack1927,Kermack1932,Kermack1933} first demonstrated how to estimate the final size of an epidemic.  Specifically, they considered a scenario such that:
%
% \begin{enumerate}
\begin{enumerate*}[label={\emph{(\roman*)}}]
  \item the disease results in complete immunity or death,
  \item all individuals are equally susceptible,
  \item the disease is transmitted in a closed population,
  \item contacts occur according to the law of mass-action,
  \item and the population is large enough to justify a deterministic analysis.
\end{enumerate*}
% \end{enumerate}
%
Under these assumptions, Kermack and McKendrick show that an epidemic with a given $R_0$ will infect a fixed fraction $R(\infty)$ of the susceptible population by solving
\begin{equation}
  R(\infty) = -\frac{1}{R_0}\ln\left[1-R(\infty)\right] \; .
  \label{eq:KMK}
\end{equation}
This solution describes a final outbreak size equal to 0 when $R_0 \leq 1$ and increasing roughly as $1-\exp(-R_0)$ when $R_0>1$.  Therefore, a larger $R_0$ leads to a larger outbreak which infects the entire population in the limit $R_0 \rightarrow \infty$. This direct relationship between $R_0$ and the final epidemic size is at the core of the conventional wisdom that a larger $R_0$ will cause a larger outbreak. % and small variations in $R_0$ can lead to vastly different total case counts. 
Unfortunately, the equation relating $R_0$ to final outbreak size from Kermack and McKendrick is only valid when all the above assumptions hold, which is rarely the case in practice. %In fact, seemingly trivial violations of the above assumptions can lead to vastly different outbreak sizes even when $R_0$ is held constant.

As a result, relying on $R_0$ alone is often misleading when comparing different pathogens or outbreaks of the same pathogen in different settings~\cite{Lloyd-Smith2005, Bansal2007, Vizi2017}.  This is especially critical considering that many outbreaks are not shaped by the ``average'' individuals but rather by a minority of super-spreading events~\cite{Lloyd-Smith2005, Meyers2005}. 
%For example, public health officials are currently trying to determine whether COVID-19 can be contained or whether it will cause a pandemic~\cite{Wu2020}. 
%Despite numerous estimates of COVID-19's $R_0$ (ranging from 1.5 to just under 4~\cite{Li2020, Read2020, Liu2020, Huang2020, Riou2020, Majumder2020}), these estimates ignore heterogeneity in the distribution of secondary infections, leading one to question if -- like severe acute respiratory syndrome (SARS) -- COVID-19 can be contained, or if -- like 2009 pandemic H1N1 influenza -- it cannot. 
%Indeed, recent work modeling the effectiveness of contact tracing and isolation on preventing COVID-19 concluded that the probability of containing the outbreak was significantly lower if COVID-19 heterogeneity in secondary infections was low, like influenza, as compared to higher, like SARS~\cite{Hellewell2020}. 
To more fully quantify how heterogeneity in the number of secondary infections affects outbreak size, we turn towards network epidemiology and derive an equation for the total number of infected individuals using all moments of the distribution of secondary infections.

\section{Random network analysis}
Random network theory allows us to relax some of assumptions made by Kermack and McKendrick, mainly to account for heterogeneity and stochasticity in the number of secondary infections caused by a given individual.  We first follow the analysis of Ref.~\cite{Newman2001} and define
\begin{equation}
  G_0(x) = \sum _{k=0}^{\infty} p_k x^k
\end{equation}
as the probability generating function (PGF) of the distribution of the number on contacts $\lbrace p_k \rbrace$ individuals have (the \textit{degree} distribution).  When following a random contact (an \textit{edge}), we define the \textit{excess} degree as the number of other edges around that node reached via one of its edges. Because an edge is $k$ times more likely to reach a node of degree $k$ than a node of degree $1$, the excess degree distribution is generated by
\begin{equation}
  G_1(x)  = \frac{G_0^\prime(x)}{G_0^\prime(1)} = \frac{1}{\langle k\rangle}\sum _{k=1}^{\infty} kp_k x^{k-1}
\end{equation}
where $\langle k \rangle$ is the average degree and acts as a normalisation constant, and $G_0^\prime(x)$ denotes the derivative of $G_0(x)$ with respect to $x$.

We now assume that the network in question is the network of all edges that \textit{would} transmit a disease if given the chance. Consequently, $G_1(x)$ generates the number of secondary infections that individual nodes would cause if infected. And, if we infect a random node as the patient zero, its entire connected component (a maximal subset of nodes between which paths exists between all pairs of nodes) will be infected. To calculate the largest possible epidemic, we thus look for the size of the giant connected component (GCC).

To calculate the size of the GCC, we first look for the probability $u$ that following a random edge leads to a node \textit{not} part of the GCC. For that node to not be part of the GCC, all of its excess edges must also not lead to the GCC. This simple observation leads to the self-consistent equation
\begin{equation} \label{eq:u}
  u = \frac{1}{\langle k\rangle}\sum _{k=1}^{\infty} kp_k u^{k-1} = G_1(u) \; .
\end{equation}
The size of the GCC is a fraction of the full population $N$ that we will denote $R({\infty})$ because it corresponds to the potential, macroscopic, outbreak size. Noting that a node of degree $k$ is \textit{not} in the GCC with probability $u^k$, $R({\infty})$ can be calculated by counting all nodes except those with no edges leading to the GCC,
\begin{equation} \label{eq:finalsize}
  R({\infty}) = 1 - G_0(u) \ .
\end{equation}
When dealing with data on the distribution of secondary infections, $G_0(x)$ has one degree of freedom remaining, $p_0$, which we set by assuring that the number of infections caused by patient zero is smaller or equal to $R_0$ but not greater (see Methods). The resulting solution for $R({\infty})$ is exact in the limit of infinite population size.

\section{Results}
The network approach naturally accounts for heterogeneity, meaning that some individuals will cause more infections than others.  The network approach also accounts for stochasticity explicitly: Even with $R_0 > 1$, there is a probability $1-R({\infty})$ that patient zero lies outside of the giant outbreak and therefore only leads to a small outbreak that does not invade the population.  However, the analysis in terms of PGFs is obviously more involved than simply assuming mass-action mixing and solving Eq.~(\ref{eq:KMK}).  In fact, the PGF $G_0(x)$ requires a full distribution of secondary cases per primary case, which will in practice involve the specification of a high-order polynomial.

To clarify and potentially simplify the approach, we propose to reformulate the classic network model in terms of the cumulant generating function (CGF) of secondary cases.  The CGF $K(y)$ of a random variable $X$ can be written as $K(y) = \sum \kappa_n y^n / n!$ where $\kappa_n$ are the cumulants of the distribution of secondary infections.  These are useful because the cumulants are easier to interpret, i.e., $\kappa_1$ is simply the average number of secondary cases $R_0$, $\kappa_2$ is the variance, $\kappa_3$ is related to the skewness and $\kappa_4$ to the kurtosis of the full distribution, etc. By definition, a PGF $G(x)$ of a random variable is linked to $K(y)$ through $G(x) = \exp\left[K\left(\ln x\right)\right]$. Therefore, we can replace the PGF $G_1(x)$ for the distribution of secondary infections by a function in terms of the cumulants of that distribution.

\begin{figure}[t]
    \centering
    \includegraphics[width=0.49\linewidth]{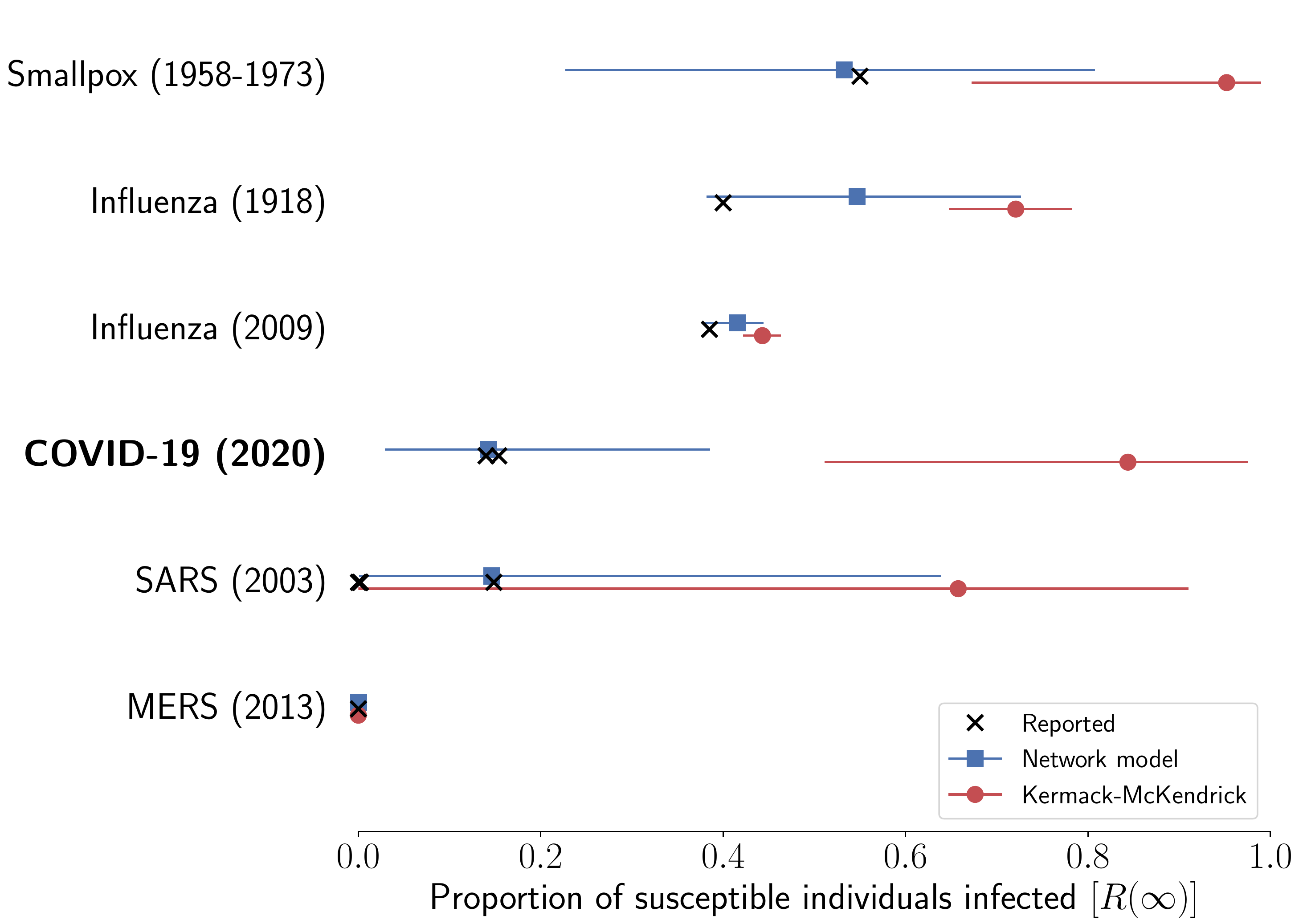}%
    \includegraphics[width=0.49\linewidth]{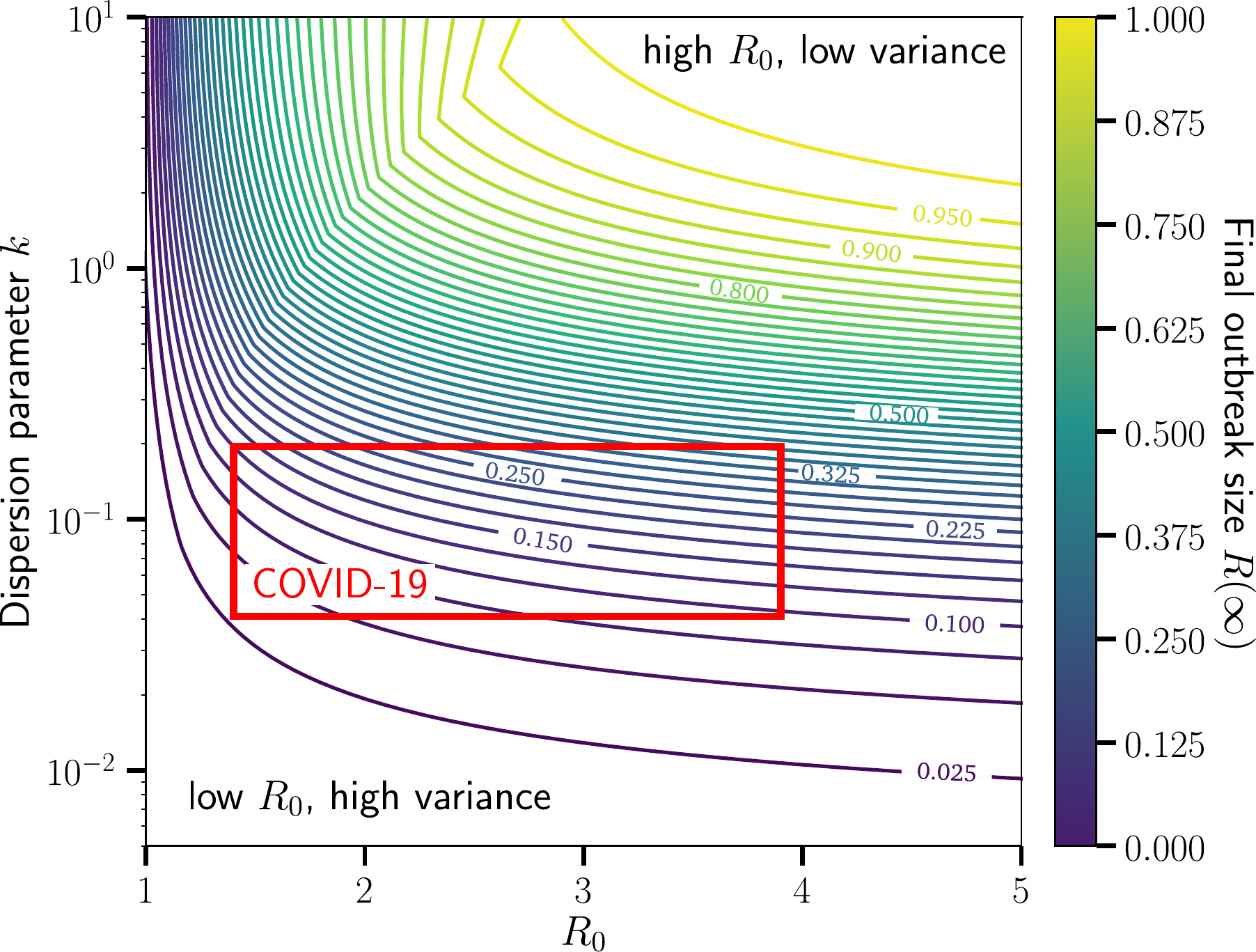}
    \caption{(left) Using published estimates of $R_0$ and the dispersion parameter, $k$, we estimated the total outbreak size for six different diseases.  The confidence intervals span the range of uncertainty reported for $R_0$ and $k$.  %Because no published estimates of $k$ exist for COVID-19, following \cite{Riou2020} we used a range of values from SARS (2003). 
    The black markers show reported total outbreak sizes (total proportion of susceptible individuals infected) for each disease.  For influenza, we report the estimated proportion of school-aged children infected.  The blue squares show the estimated proportion infected obtained with Eq.~\eqref{eq:finalsize}.  The red circles are the estimated proportion infected using the method developed by Kermack and McKendrick, i.e., Eq.~\eqref{eq:KMK}.  
    %The green triangles are the estimated proportion infected using the approximation $1-\frac{1}{R_0}$.  
    % See Table~\ref{tab:propI} and \url{https://github.com/Emergent-Epidemics/beyond_R0} for additional information and to access the data. % Ideally, validation should use estimates of final outbreak size, $R_0$ and $k$ inferred from the same population, but unfortunately that is rarely, if ever, available.
    (right) Final size of outbreaks with different $R_0$ and distributions of secondary cases. We use a negative binomial distribution of secondary cases and scan a realistic range of parameters.  The range of parameters corresponding to current estimates for COVID-19 is highlighted by a red box.
    %The range of potential $R_0$ comes from a 95\% confidence interval using a early data and a classic deterministic models \cite{Li2020}. The range of dispersion parameter $k$ comes from analogy with SARS (2003) \cite{Lloyd-Smith2005, Hellewell2020}.
    Most importantly, with fixed average, the dispersion parameter is inversely proportional to the variance of the underlying distribution of second cases.
    See Table~\ref{tab:propI} and \url{https://github.com/Emergent-Epidemics/beyond_R0} for additional information and to access the data.
    }
    \label{fig:diseases}
\end{figure}

\subsection{Analysis of cumulants and derivation of Kermack-McKendrick}
%
%Using the cumulants $\kappa _n$, we can re-write $G_1(x)$ as
%%
%\begin{align}
%  G^C_1(x) = \exp \left[\sum_{n=1}^\infty \frac{1}{n!}\kappa_n \left(\ln x\right)^n \right] = G_1(x) \; .
%  \label{GC1}
%\end{align}
%
%which while not as practical as a simple polynomial when solving for $u=G^C_1(u)$, in some cases, cumulants can have significant advantage over the polynomial representation. For instance, the cumulants of a sum of independent random variables are simply the sum of the cumulants --- convenient when dealing with diseases with multiple modes of transmission.
%\begin{figure}[t]
%  \centering
%  \includegraphics[width=\linewidth]{fig_same_R0_different_Rinf.pdf}
%  \caption{Illustration of the absence of a direct relationship between $R_0$ and the final outbreak size, $R(\infty)$, using three different distributions for the number of secondary infections per infected individual. All three distributions are negative binomials~\cite{Lloyd-Smith2005} with distinct average, $R_0$, and variance, $\sigma^2$.  $R(\infty)$ is computed by solving Eqs.~\eqref{eq:u}~and~\eqref{eq:finalsize}.}
%  \label{fig:same_R0_different_Rinf}
%\end{figure}
We can easily derive Kermack and McKendrick's result from this framework since their solution assumes a well-mixed population, which corresponds to a Poisson distribution of secondary infections.  We first re-write $G_1(x)$ in terms of the cumulants $\kappa _n$ as%
\begin{equation}
  G_1(x) = \exp \left[\sum_{n=1}^\infty \frac{1}{n!}\kappa_n \left(\ln x\right)^n \right] \ ,
  \label{GC1}
\end{equation}
which is a particular convenient representation for a Poisson distribution because its cumulants $\kappa_n = R_0$ for all $n > 0$.  Moreover, since $G_0(x) = G_1(x)$ in the Poisson case, the final outbreak size of the Kermack-McKendrick analysis will be set by $u_{\textrm{KM}} = G_1(u_{\textrm{KM}})$, or
\begin{eqnarray}
  u_{\textrm{KM}} & = & \exp\left[ \sum_{n=1}^ \infty \frac{1}{n!}R_0\left(\ln u_{\textrm{KM}}\right)^n\right] =  \exp \Big[R_0\left(u_{\textrm{KM}}-1\right)\Big] \ ; \nonumber \\
  \hookrightarrow\ & & R_{\textrm{KM}}(\infty) =  1 - \exp \left[R_0\left(u_{\textrm{KM}}-1\right)\right] =  1 - \exp \left[-R_0 R_{\textrm{KM}}(\infty)\right]
\end{eqnarray}
Taking the logarithm of the exponential term from this last equation yields Kermack and McKendrick's formula.

The solution to $u = G_1(u)$ gives the probability that every infection caused by patient zero fails to generate an epidemic.  For more general distributions, it is useful to rewrite Eq.~(\ref{GC1}) as
\begin{eqnarray} \label{GC2}
  u = G_1(u) & = & \exp \left[\sum_{n=1}^\infty \frac{1}{n!}\kappa_n \left(\ln u\right)^n \right] \\
  & = & \exp \left[R_0 \vert \ln u \vert - \frac{1}{2}\sigma^2 \vert \ln u \vert^2 + \frac{1}{6}\kappa_3 \vert \ln u \vert^3 - \frac{1}{24}\kappa_4 \vert \ln u \vert^4 \ldots\right] \nonumber
\end{eqnarray}
to highlight its alternating nature because $\ln u$ is negative ($u$ is a probability) such that its $n$-th power is positive when $n$ is even and negative when $n$ is odd.

The alternating sign of contribution from high-order moments in Eq. (\ref{GC2}) can be interpreted as follows. A disease needs a high average number of secondary infections (high $\kappa_1 = R_0$) to spread, but given that average, a disease with small variance in secondary infections will spread much more reliably and be less likely to stochastically die out. % (see Fig.~\ref{fig:same_R0_different_Rinf}). 
Given a variance, a disease with high skewness (i.e., with positive deviation contributing to most of the variance) will be more stable than a disease with negative skewness (i.e. with most deviations being towards small secondary infections). Given a skewness, a disease will be more stable if it has frequent small positive deviations rather than infrequent large deviations --- hence a smaller kurtosis --- as stochastic die out could easily occur before any of those large infrequent deviations occur.

Our re-interpretation already highlights a striking result: Higher moments of the distribution of secondary cases can lead a disease with a lower $R_0$ to invade more easily a population and to reach a larger final outbreak size than a disease with a higher $R_0$.  Taking into account the contribution of these higher moments also yields better estimates for the final size of outbreaks, as we now show.
% We can investigate this conclusion further using a simple example of normally distributed secondary infections.

\begin{table}[t]
  %\footnotesize
  \centering
  \begin{tabular}{llccccc}
    \hline
    Disease   & Location        & Year      & Prop. Infect. & $R_0$ & $k$ & Ref. \\ 
    \hline
    MERS      & Global          & 2013      & 0.00          & 0.47 (0.29--0.80)\textsuperscript{\S}      & 0.26 (0.09--1.24)\textsuperscript{\S}      & \cite{Memish2014,Kucharski2015} \\ 
    SARS      & Global          & 2003      & 0.00--0.15    & 1.63 (0.54--2.65)\textsuperscript{$\star$} & 0.16 (0.11--0.64)\textsuperscript{$\star$} & \cite{leung_seroprevalence_2006,Quah2004,Lloyd-Smith2005} \\ 
    Smallpox  & Europe          & 1958-1973 & 0.55          & 3.19 (1.66--4.62)\textsuperscript{$\star$} & 0.37 (0.26--0.69)\textsuperscript{$\star$} & \cite{Mack1972,Lloyd-Smith2005} \\ 
    Influenza & Baltimore (USA) & 1918      & 0.40          & 1.77 (1.61--1.95)\textsuperscript{\S}      & 0.94 (0.59--1.72)\textsuperscript{\S}      & \cite{Taubenberger2006,Fraser2011} \\ 
    Influenza & Italy           & 2009      & 0.39          & 1.321 (1.299--1.343)\textsuperscript{\S}   & 8.092 (5.170--11.794)\textsuperscript{\S}  & \cite{Kelly2011,Dorigatti2012} \\
    COVID-19  & Global          & 2020      & 0.14, 0.154    & 2.5 (1.4--12)\textsuperscript{\S}          & 0.1 (0.04--0.2)\textsuperscript{\S} & \cite{Li2020,endo_estimating_2020,Streeck2020,sutton_universal_2020} \\
%    COVID-19  & China           & 2020      & ---           & 2.20 (1.4--3.8)\textsuperscript{$\star$}   & 0.20 (0.11--0.64)\textsuperscript{$\star$} & \cite{Li2020,Read2020,Lloyd-Smith2005,Hellewell2020} \\
    \hline
  \end{tabular}
  \caption{Estimates for $R_0$ and for the negative binomial distribution dispersion parameter, $k$, used on Fig.~\ref{fig:diseases} (\S\ and $\star$ respectively denote 95\% and 90\% confidence intervals).  Proportion of susceptible individuals infected as reported in either the literature or by the US CDC.  For SARS, the proportion of infected was taken from serosurveys among wild animal handlers (0.15) and among health-care workers ($<$0.01)~\cite{leung_seroprevalence_2006}.  For influenza (2009), we took data on school-aged children.  For COVID-19, the proportion of infected was taken from a serosurvey in the municipality of Gangelt, Germany~\cite{Streeck2020} and from universal testing in all obstetrical patients presenting for delivery at two hospitals~\cite{sutton_universal_2020}.  Note that the estimates the proportion of infected individuals, for $R_0$ and for $k$ were not necessarily inferred from the same populations.   Such information is rarely, if ever, available for a same outbreak, unfortunately.
  % Ideally, validation should use estimates of final outbreak size, $R_0$ and $k$ inferred from the same population, but unfortunately that is rarely, if ever, available.
  % Estimates for $R_0$ and for the negative binomial distribution dispersion parameter, $k$, as reported in the literature.  
  % As of writing there were no published estimates of $k$ for COVID-19, we used a range for $k$ values from SARS (2003).
  % 95\% and 90\% confidence intervals are denoted by \S\ and $\star$ respectively.
  }
  \label{tab:propI}
\end{table}

%\subsection{Normal distributions and the impact of variance}
%
%A second useful application of the cumulants formulation involves diseases with a large reproductive number $R_0$ whose distribution of secondary infections can be convincingly modeled by a normal distribution. The raw moments of a normal distribution are quite complicated, but the cumulants are simple: $\kappa_1$ is equal to the mean $R_0$, $\kappa_2$ is equal to the variance $\sigma^2$, and all other cumulants are 0. We can thus write

%\begin{equation}
%  G^C_1(x) = \exp \left[ R_0 \ln x + \frac{1}{2}\sigma^2 \left(\ln x\right)^2 \right] = x^{R_0 + \frac{\sigma^2}{2}\ln x}
%\end{equation}
%and solving for $u=G^C_1(u)$ yields
%\begin{equation}
%  u = \exp \left[-\frac{2}{\sigma^2}\left(R_0-1\right)\right] \; .
%\end{equation}

%This equation can then be used for direct comparison of the probability of invasion of two different diseases with normal distributions of secondary infections. Given a transmission event from patient zero to a susceptible individual, disease B will be more likely to invade the population than disease A if

%\begin{equation}
%  \frac{\sigma_A^2}{\sigma_B^2} < \frac{R_{0,A}-1}{R_{0,B}-1} \; .
%  \label{eq:ineq}
%\end{equation}

%For example, a disease with half the basic reproductive number of another will still be more likely to invade a population and lead to a larger outbreak if its variance is less or close to half the variance of the other disease.

\subsection{Comparison of estimators to empirical data}
We now compare the final outbreak size estimates from Eq.~\eqref{eq:KMK} (Kermack and McKendrick) to estimates from Eq.~\eqref{eq:finalsize} (network model) with a negative binomial offspring distribution (see Methods and Table~\ref{tab:propI}).  As predicted, Fig.~\ref{fig:diseases}(left) shows that the Kermack and McKendrick formulation consistently and significantly over-predicts the outbreak size across six different pathogens where we could find confidence interval estimates for $R_0$ and for the negative binomial over-dispersion parameter ($k$). %, we find that -- as predicted -- the Kermack and McKendrick formulation consistently and significantly over-predicts the outbreak size, see Fig.~\ref{fig:diseases}.  
%Conversely, for diseases with low $R_0$ and a high value for $k$ like the 2009 H1N1 influenza pandemic, indicating reduced variability in secondary infections, the ``herd immunity'' approximation under predicts the outbreak size. 
Our approach produces estimates of the total outbreak size, which are consistent with outbreaks where no vaccine was available (smallpox in unvaccinated populations, the 1918 influenza pandemic, and school children prior to the availability of the 2009 H1N1 vaccine). Clearly, once interventions are put in place and\slash or substantial behavioral change occurs, all methods that do not account for these effects will over-estimate the total outbreak size~\cite{Eksin2019}.  Nevertheless, our approach provides a much more reasoned estimate of the total risk to any given population, and predictions very close to the most recent seropositivity estimates for the COVID-19 outbreak in a German Municipality~\cite{Streeck2020} and in obstetrical patients presenting for delivery~\cite{sutton_universal_2020},  as well as for SARS among wild animal handlers (other smaller estimates correspond to health-care workers)~\cite{leung_seroprevalence_2006}.
%Ideally, validation should use estimates of final outbreak size, $R_0$ and $k$ inferred from the same population, but unfortunately that is rarely, if ever, available.

%\begin{figure}
%    \centering
%    \includegraphics[width=0.6\linewidth]{contour_newest_test.pdf}
%    \caption{Final size of outbreaks with different $R_0$ and distributions of secondary cases. We use a negative binomial distribution of secondary cases and scan a realistic range of parameters.  The range of parameters corresponding to current estimates for COVID-19 (see Table~\ref{tab:propI}) is highlighted by a red box.
%    %The range of potential $R_0$ comes from a 95\% confidence interval using a early data and a classic deterministic models \cite{Li2020}. The range of dispersion parameter $k$ comes from analogy with SARS (2003) \cite{Lloyd-Smith2005, Hellewell2020}.
%    Most importantly, with fixed average, the dispersion parameter is inversely proportional to the variance of the underlying distribution of second cases.}
%    \label{fig:ncov}
%\end{figure}

\section{Discussion}
From re-emerging pathogens like yellow fever and measles to emerging threats like Middle East Respiratory Syndrome coronavirus and Ebola, the World Health Organization monitored 119 different infectious disease outbreaks in 2019 alone~\cite{WHO19}. For each of these outbreaks, predicting both the epidemic potential and the most likely number of cases is critically important for efficient and effective responses. This need for rapid situational awareness is why $R_0$ is so widely used in public health. However, our main analysis shows that not only is $R_0$ insufficient in fully determining the final size of an outbreak, but having a larger outbreak with a lower $R_0$ is relatively easy considering the randomness associated with most transmission events and the heterogeneity of physical contacts. To address the need for rapid quantification of risk, while acknowledging the shortcomings of $R_0$, we use network science methods to derive both the probability of an epidemic and its final size.

%removing references in 2nd block: Watts2005, 3rd block: Colizza2006a, 4th block Althouse2014, and removed this clause: higher-order contact structure~\cite{Hebert-Dufresne2010, Volz2011}
These results are not without important caveats.  Specifically, we must remember that distributions of secondary cases, just like $R_0$ itself, are just as much a product of a pathogen as of the population in which it spreads. For example, aspects of the social contact network~\cite{Moreno2002}, metapopulation structure~\cite{Colizza2008}, human mobility~\cite{Wesolowski2015}, adaptive behavior~\cite{Scarpino2016}, and even other pathogens~\cite{Hebert-Dufresne2015, Hebert-Dufresne2019}, all interact to cause complex patterns of disease emergence, spread, and persistence. Therefore, great care must be taken when using any of these tools to compare outbreaks or to inform current events with past data.

Figure \ref{fig:diseases}(left) only used a handful of known outbreaks to validate the different approaches because data on secondary cases are rare.  In practice, three types of data could potentially be used in real time to improve predictions by considering secondary case heterogeneity. First, contact tracing data whose objective is to identify people who may have come into contact with an infectious individual. While mostly a preventive measure to identify cases before complications, it directly informs us about potential secondary cases caused by a single individual, and therefore provides us with an estimate for $G_1(x)$.  Both for generating accurate predictions of epidemic risk and controlling the outbreak, it is vital to begin contact tracing before numerous transmission chains become widely distributed across space~\cite{Dhillon2018, Klinkenberg2006}.

%removing some references in 1st block: Gire2014, Drosten2015, 2nd block: Arias2016, Park2015, 4th block: Stadler2013
Second, viral genome sequences provide information on both the timing of the outbreak~\cite{Smith2009} and structure of secondary cases~\cite{Scarpino2015}.  For example, methods exist to reconstruct transmission trees for sampled sequences using simple mutational models to construct a likelihood for a specific transmission tree~\cite{Jombart2014, Campbell2019} and translate coalescent rates into key epidemiological parameters~\cite{Volz2013, Bouckaert2019}. Despite the potential for genome sequencing to revolutionize outbreak response, the global public health community still struggles to coordinate data sharing across international borders, between academic researchers, and with private companies~\cite{Gardy2015, VanPuyvelde2019, Grubaugh2019}.

%Third, and most often the first available information on novel pathogens, are data related to similar past outbreaks. For example, in Figs.~\ref{fig:diseases} and \ref{fig:ncov} we make predictions for the final size of COVID-19 based on $R_0$ estimates from early cases  \cite{Li2020,Read2020} and for the underlying distribution of secondary cases by analogy with the SARS (2003) in similar population \cite{Lloyd-Smith2005}. Based on this uncertainty, we obtain a range of final outbreak size (given as a fraction of the total susceptible population) between 10\% and 75\%. Given the large difference seen in Fig.~\ref{fig:diseases} between our prediction for COVID-19 and that of mass-action frameworks, there is a dire need for contact tracing data and\slash high-resolution pathogen genome surveillance for this and any future emerging outbreak so that heterogeneity can be estimated accurately.

%\begin{figure}
%    \centering
%    \includegraphics[width=0.8\linewidth]{histogram_BMA.pdf}
%    \caption{Mapping a posterior distribution of basic reproduction number $R_0$ and dispersion parameter $k$, based on \textit{very early} data for the COVID-19 outbreak as available on January 24th 2020 \cite{Riou2020}, to a posterior distribution of final sizes. Unlike the analysis of Fig.~\ref{fig:ncov} which borrowed parameters from related diseases, this analysis leverages stochastic simulations and early incidence data to provide a probabilistic forecast of different outbreak sizes.}
%    \label{fig:posterior}
%\end{figure}

Third, early incidence data can be leveraged to infer parameters of the secondary case distribution through comparison with simulations. Comparing the output of agent-based simulations with reported incidence can be used to effectively sample a joint posterior distribution over $R_0$ and dispersion parameter $k$. This approach was used by most studies referenced in Table~\ref{tab:propI}.  Most importantly, these simulations need not be run over long periods of time to predict final outbreak size.  Instead, they only need to be run over enough early data to infer the parameter estimates that are then fed into our network model to compute the final outbreak size.

As for COVID-19, Fig.~\ref{fig:diseases}(right) shows how the width of the confidence interval on our prediction for the final outbreak size mostly stems from uncertainty in the heterogeneity of secondary infections; i.e., the dispersion parameter $k$. With limited heterogeneity, our predictions would have been closer to classic mass-action forecasts and the current pandemic of COVID-19 would likely have been a consequence of not only $R_0$, but of the homogeneity of secondary infections: each new cases steadily leading to additional infections. Thankfully, with recent large estimates for its heterogeneity, the observed transmission could be mostly maintained by so-called ``super-spreading events", which could be easier to manage with contact tracing, screening and infection control~\cite{Hellewell2020}.

In conclusion, we reiterate that when accounting for the full distribution of secondary cases caused by an infected individual, there is no direct relationship between $R_0$ and the size of an outbreak. 
We also stress that both $R_0$ and the full secondary case distribution are not properties of the disease itself, but are instead set by properties of the pathogen, the host population and the context of the outbreak. 
Nevertheless, we provide a straightforward methodology for translating estimates of transmission heterogeneity into epidemic forecasts. 
Altogether, predicting outbreak size based on early data is an incredibly complex challenge but one that is increasingly within reach due to new mathematical analyses and faster communication of public health data.

\section*{Acknowledgments}
L.H.-D. acknowledges support from the National Institutes of Health 1P20 GM125498-01 Centers of Biomedical Research Excellence Award. B.M.A. is supported by Bill and Melinda Gates through the Global Good Fund. S.V.S. is supported by startup funds provided by Northeastern University. A.A. acknowledges financial support from the Sentinelle Nord initiative of the Canada First Research Excellence Fund and from the Natural Sciences and Engineering Research Council of Canada (project 2019-05183).

\section*{Methods}
The results presented from our network model assume the number of secondary infections to be distributed according to a negative binomial distribution parametrized by its average $R_0$ and dispersion $k$~\cite{Lloyd-Smith2005}. Its probability generating function (PGF) is
\begin{equation}
  G_1(x)
    = \sum_{n=0}^{\infty} \binom{n+k-1}{n} \left[ \frac{R_0}{R_0 + k} \right]^{n} \left[ 1 - \frac{R_0}{R_0 + k} \right]^{k} x^n
    = \left[ 1 + \frac{R_0}{k}(1 - x) \right]^{-k} \ .
\end{equation}

The network theory formalism presented in the main text requires the specification of the PGF $G_0(x)$ whose related to $G_1(x)$ via
\begin{equation} \label{eq:G1_definition}
  G_1(x) = \frac{G_0^\prime(x)}{G_0^\prime(1)}
\end{equation}
where the prime ($\prime$) denotes the first derivative with respect to $x$.  Specifying $G_1(x)$ therefore fixes $G_0(x)$ up to a constant and to a multiplicative factor.  Without loss of generality, we set
\begin{equation}
  G_0(x) = p_0 + (1 - p_0) g_0(x)
\end{equation}
with $0 \leq p_0 \leq 1$, $g_0(0) = 0$ and $g_1(1) = 1$.  Equation~\eqref{eq:G1_definition} becomes
\begin{equation}
  G_1(x) = \frac{g_0^\prime(x)}{g_0^\prime(1)} \ ,
\end{equation}
from which we compute
\begin{equation}
  g_0(x)
    = \int g_0^\prime(x) dx
    = g_0^\prime(1) \int G_1(x) dx
    = -\frac{k g_0^\prime(1)}{R_0 (1 - k)} \left[ 1 + \frac{R_0}{k}(1 - x) \right]^{1-k} + C \ ,
\end{equation}
with $k \neq 1$, and  where $C$ and $g_0^\prime(1)$ are fixed by imposing $g_0(0) = 0$ and $g_1(1) = 1$.  Rearranging the terms, we find that
\begin{equation}
  g_0(x)
    = \frac{1 - \left[ 1 - \frac{R_0 x}{R_0 + k} \right]^{1-k}}{1 - \left[ \frac{k}{R_0 + k} \right]^{1-k}} \ ,
\end{equation}
from which we finally obtain
\begin{equation} \label{eq:G0_k_neq_1}
  G_0(x) = p_0 + (1 - p_0) \frac{1 - \left[ 1 - \frac{R_0 x}{R_0 + k} \right]^{1-k}}{1 - \left[ \frac{k}{R_0 + k} \right]^{1-k}}
\end{equation}
with $k \neq 1$.  The case $k=1$ must be treated separately and yields
\begin{equation} \label{eq:G0_k_eq_1}
  G_0(x) = p_0 + (1 - p_0) \left[ 1 - \frac{\ln \left[ 1 + R_0 (1 - x) \right]}{\ln \left[ 1 + R_0 \right]} \right] \ .
\end{equation}

From Eqs.~\eqref{eq:G0_k_neq_1}~and~\eqref{eq:G0_k_eq_1}, we find that the average number of secondary infections caused by the patient zero is
\begin{equation}
  G_0^\prime(1) = (1 - p_0)\frac{(1 - k)R_0}{k} \frac{1}{\left[ \frac{k}{R_0 + k} \right]^{k-1} - 1}
\end{equation}
if $k \neq 1$, and
\begin{equation}
  G_0^\prime(1) = (1 - p_0) \frac{R_0}{\ln \left[1 + R_0 \right]}
\end{equation}
if $k = 1$.  The average number of secondary infections caused by patient zero can therefore be greater or smaller than $R_0$. Since patient zero should not be expected to create \textit{more} secondary cases than the next generation of infections, we set the value of $p_0 \in [0, 1]$ such that $G_0^\prime(1)$ is as close as possible to $R_0$ whenever $G_0^\prime(1) > R_0$.
%Throughout the main text, we consider the case $p_0 = 0$, which can be interpreted as assuming that we only consider susceptible individuals (i.e., individuals that \textit{could} get infected).

A large-scale epidemic is predicted by this framework~\cite{Newman2001} if
\begin{equation}
  G_1^\prime(1) = R_0 > 1 \ ,
\end{equation}
as in the analysis by Kermack and McKendrick~\cite{Kermack1927,Kermack1932,Kermack1933}.  Its size, $R(\infty)$, is computed with $G_0(x)$ as
\begin{equation}
  R(\infty) = 1 - G_0(u)
\end{equation}
where $u$ is the solution of
\begin{equation}
  u = G_1(u)
\end{equation}
which we solve using the relaxation method~\cite{Newman2012b} with an initial condition randomly chosen in the open interval $(0,1)$.


\begin{thebibliography}{50}%
\makeatletter
\providecommand \@ifxundefined [1]{%
 \@ifx{#1\undefined}
}%
\providecommand \@ifnum [1]{%
 \ifnum #1\expandafter \@firstoftwo
 \else \expandafter \@secondoftwo
 \fi
}%
\providecommand \@ifx [1]{%
 \ifx #1\expandafter \@firstoftwo
 \else \expandafter \@secondoftwo
 \fi
}%
\providecommand \natexlab [1]{#1}%
\providecommand \enquote  [1]{``#1''}%
\providecommand \bibnamefont  [1]{#1}%
\providecommand \bibfnamefont [1]{#1}%
\providecommand \citenamefont [1]{#1}%
\providecommand \href@noop [0]{\@secondoftwo}%
\providecommand \href [0]{\begingroup \@sanitize@url \@href}%
\providecommand \@href[1]{\@@startlink{#1}\@@href}%
\providecommand \@@href[1]{\endgroup#1\@@endlink}%
\providecommand \@sanitize@url [0]{\catcode `\\12\catcode `\$12\catcode
  `\&12\catcode `\#12\catcode `\^12\catcode `\_12\catcode `\%12\relax}%
\providecommand \@@startlink[1]{}%
\providecommand \@@endlink[0]{}%
\providecommand \url  [0]{\begingroup\@sanitize@url \@url }%
\providecommand \@url [1]{\endgroup\@href {#1}{\urlprefix }}%
\providecommand \urlprefix  [0]{URL }%
\providecommand \Eprint [0]{\href }%
\providecommand \doibase [0]{http://dx.doi.org/}%
\providecommand \selectlanguage [0]{\@gobble}%
\providecommand \bibinfo  [0]{\@secondoftwo}%
\providecommand \bibfield  [0]{\@secondoftwo}%
\providecommand \translation [1]{[#1]}%
\providecommand \BibitemOpen [0]{}%
\providecommand \bibitemStop [0]{}%
\providecommand \bibitemNoStop [0]{.\EOS\space}%
\providecommand \EOS [0]{\spacefactor3000\relax}%
\providecommand \BibitemShut  [1]{\csname bibitem#1\endcsname}%
\let\auto@bib@innerbib\@empty
%</preamble>
\bibitem [{\citenamefont {Lloyd-Smith}\ \emph {et~al.}(2005)\citenamefont
  {Lloyd-Smith}, \citenamefont {Schreiber}, \citenamefont {Kopp},\ and\
  \citenamefont {Getz}}]{Lloyd-Smith2005}%
  \BibitemOpen
  \bibfield  {author} {\bibinfo {author} {\bibfnamefont {J.~O.}\ \bibnamefont
  {Lloyd-Smith}}, \bibinfo {author} {\bibfnamefont {S.~J.}\ \bibnamefont
  {Schreiber}}, \bibinfo {author} {\bibfnamefont {P.~E.}\ \bibnamefont {Kopp}},
  \ and\ \bibinfo {author} {\bibfnamefont {W.~M.}\ \bibnamefont {Getz}},\
  }\bibfield  {title} {\enquote {\bibinfo {title} {{Superspreading and the
  effect of individual variation on disease emergence}},}\ }\href {\doibase
  10.1038/nature04153} {\bibfield  {journal} {\bibinfo  {journal} {Nature}\
  }\textbf {\bibinfo {volume} {438}},\ \bibinfo {pages} {355--359} (\bibinfo
  {year} {2005})}\BibitemShut {NoStop}%
\bibitem [{\citenamefont {Bansal}\ \emph {et~al.}(2007)\citenamefont {Bansal},
  \citenamefont {Grenfell},\ and\ \citenamefont {Meyers}}]{Bansal2007}%
  \BibitemOpen
  \bibfield  {author} {\bibinfo {author} {\bibfnamefont {S.}~\bibnamefont
  {Bansal}}, \bibinfo {author} {\bibfnamefont {B.~T.}\ \bibnamefont
  {Grenfell}}, \ and\ \bibinfo {author} {\bibfnamefont {L.~A.}\ \bibnamefont
  {Meyers}},\ }\bibfield  {title} {\enquote {\bibinfo {title} {{When individual
  behaviour matters: homogeneous and network models in epidemiology}},}\ }\href
  {\doibase 10.1098/rsif.2007.1100} {\bibfield  {journal} {\bibinfo  {journal}
  {J. R. Soc. Interface}\ }\textbf {\bibinfo {volume} {4}},\ \bibinfo {pages}
  {879--891} (\bibinfo {year} {2007})}\BibitemShut {NoStop}%
\bibitem [{\citenamefont {Vizi}\ \emph {et~al.}(2017)\citenamefont {Vizi},
  \citenamefont {Kiss}, \citenamefont {Miller},\ and\ \citenamefont
  {R{\"{o}}st}}]{Vizi2017}%
  \BibitemOpen
  \bibfield  {author} {\bibinfo {author} {\bibfnamefont {Z.}~\bibnamefont
  {Vizi}}, \bibinfo {author} {\bibfnamefont {I.~Z.}\ \bibnamefont {Kiss}},
  \bibinfo {author} {\bibfnamefont {J.~C.}\ \bibnamefont {Miller}}, \ and\
  \bibinfo {author} {\bibfnamefont {G.}~\bibnamefont {R{\"{o}}st}},\ }\bibfield
   {title} {\enquote {\bibinfo {title} {{A monotonic relationship between the
  variability of the infectious period and final size in pairwise epidemic
  modelling}},}\ }\href@noop {} {\bibfield  {journal} {\bibinfo  {journal}
  {arXiv}\ } (\bibinfo {year} {2017})},\ \Eprint
  {http://arxiv.org/abs/1712.06026} {arXiv:1712.06026} \BibitemShut {NoStop}%
\bibitem [{\citenamefont {Newman}\ \emph {et~al.}(2001)\citenamefont {Newman},
  \citenamefont {Strogatz},\ and\ \citenamefont {Watts}}]{Newman2001}%
  \BibitemOpen
  \bibfield  {author} {\bibinfo {author} {\bibfnamefont {M.~E.~J.}\
  \bibnamefont {Newman}}, \bibinfo {author} {\bibfnamefont {S.~H.}\
  \bibnamefont {Strogatz}}, \ and\ \bibinfo {author} {\bibfnamefont {D.~J.}\
  \bibnamefont {Watts}},\ }\bibfield  {title} {\enquote {\bibinfo {title}
  {{Random graphs with arbitrary degree distributions and their
  applications}},}\ }\href {\doibase 10.1103/PhysRevE.64.026118} {\bibfield
  {journal} {\bibinfo  {journal} {Phys. Rev. E}\ }\textbf {\bibinfo {volume}
  {64}},\ \bibinfo {pages} {026118} (\bibinfo {year} {2001})}\BibitemShut
  {NoStop}%
\bibitem [{\citenamefont {Biggerstaff}\ \emph {et~al.}(2014)\citenamefont
  {Biggerstaff}, \citenamefont {Cauchemez}, \citenamefont {Reed}, \citenamefont
  {Gambhir},\ and\ \citenamefont {Finelli}}]{Biggerstaff2014}%
  \BibitemOpen
  \bibfield  {author} {\bibinfo {author} {\bibfnamefont {M.}~\bibnamefont
  {Biggerstaff}}, \bibinfo {author} {\bibfnamefont {S.}~\bibnamefont
  {Cauchemez}}, \bibinfo {author} {\bibfnamefont {C.}~\bibnamefont {Reed}},
  \bibinfo {author} {\bibfnamefont {M.}~\bibnamefont {Gambhir}}, \ and\
  \bibinfo {author} {\bibfnamefont {L.}~\bibnamefont {Finelli}},\ }\bibfield
  {title} {\enquote {\bibinfo {title} {{Estimates of the reproduction number
  for seasonal, pandemic, and zoonotic influenza: a systematic review of the
  literature}},}\ }\href {\doibase 10.1186/1471-2334-14-480} {\bibfield
  {journal} {\bibinfo  {journal} {BMC Infect. Dis.}\ }\textbf {\bibinfo
  {volume} {14}},\ \bibinfo {pages} {480} (\bibinfo {year} {2014})}\BibitemShut
  {NoStop}%
\bibitem [{\citenamefont {{WHO Ebola Response
  Team}}(2014)}]{WHOEbolaResponseTeam2014}%
  \BibitemOpen
  \bibfield  {author} {\bibinfo {author} {\bibnamefont {{WHO Ebola Response
  Team}}},\ }\bibfield  {title} {\enquote {\bibinfo {title} {{Ebola Virus
  Disease in West Africa — The First 9 Months of the Epidemic and Forward
  Projections}},}\ }\href {\doibase 10.1056/NEJMoa1411100} {\bibfield
  {journal} {\bibinfo  {journal} {N. Engl. J. Med.}\ }\textbf {\bibinfo
  {volume} {371}},\ \bibinfo {pages} {1481--1495} (\bibinfo {year}
  {2014})}\BibitemShut {NoStop}%
\bibitem [{\citenamefont {Althaus}(2014)}]{Althaus2014}%
  \BibitemOpen
  \bibfield  {author} {\bibinfo {author} {\bibfnamefont {C.~L.}\ \bibnamefont
  {Althaus}},\ }\bibfield  {title} {\enquote {\bibinfo {title} {{Estimating the
  Reproduction Number of Ebola Virus (EBOV) During the 2014 Outbreak in West
  Africa}},}\ }\href {\doibase
  10.1371/currents.outbreaks.91afb5e0f279e7f29e7056095255b288} {\bibfield
  {journal} {\bibinfo  {journal} {PLOS Curr.}\ } (\bibinfo {year} {2014}),\
  10.1371/currents.outbreaks.91afb5e0f279e7f29e7056095255b288}\BibitemShut
  {NoStop}%
\bibitem [{\citenamefont {Mills}\ \emph {et~al.}(2004)\citenamefont {Mills},
  \citenamefont {Robins},\ and\ \citenamefont {Lipsitch}}]{Mills2004}%
  \BibitemOpen
  \bibfield  {author} {\bibinfo {author} {\bibfnamefont {C.~E.}\ \bibnamefont
  {Mills}}, \bibinfo {author} {\bibfnamefont {J.~M.}\ \bibnamefont {Robins}}, \
  and\ \bibinfo {author} {\bibfnamefont {M.}~\bibnamefont {Lipsitch}},\
  }\bibfield  {title} {\enquote {\bibinfo {title} {{Transmissibility of 1918
  pandemic influenza}},}\ }\href {\doibase 10.1038/nature03063} {\bibfield
  {journal} {\bibinfo  {journal} {Nature}\ }\textbf {\bibinfo {volume} {432}},\
  \bibinfo {pages} {904--906} (\bibinfo {year} {2004})}\BibitemShut {NoStop}%
\bibitem [{\citenamefont {Kaner}\ and\ \citenamefont
  {Schaack}(2016)}]{Kaner2016}%
  \BibitemOpen
  \bibfield  {author} {\bibinfo {author} {\bibfnamefont {J.}~\bibnamefont
  {Kaner}}\ and\ \bibinfo {author} {\bibfnamefont {S.}~\bibnamefont
  {Schaack}},\ }\bibfield  {title} {\enquote {\bibinfo {title} {{Understanding
  Ebola: the 2014 epidemic}},}\ }\href {\doibase 10.1186/s12992-016-0194-4}
  {\bibfield  {journal} {\bibinfo  {journal} {Global. Health}\ }\textbf
  {\bibinfo {volume} {12}},\ \bibinfo {pages} {53} (\bibinfo {year}
  {2016})}\BibitemShut {NoStop}%
\bibitem [{\citenamefont {Bootsma}\ and\ \citenamefont
  {Ferguson}(2007)}]{Bootsma2007}%
  \BibitemOpen
  \bibfield  {author} {\bibinfo {author} {\bibfnamefont {M.~C.~J.}\
  \bibnamefont {Bootsma}}\ and\ \bibinfo {author} {\bibfnamefont {N.~M.}\
  \bibnamefont {Ferguson}},\ }\bibfield  {title} {\enquote {\bibinfo {title}
  {{The effect of public health measures on the 1918 influenza pandemic in U.S.
  cities}},}\ }\href {\doibase 10.1073/pnas.0611071104} {\bibfield  {journal}
  {\bibinfo  {journal} {Proc. Natl. Acad. Sci. USA}\ }\textbf {\bibinfo
  {volume} {104}},\ \bibinfo {pages} {7588--7593} (\bibinfo {year}
  {2007})}\BibitemShut {NoStop}%
\bibitem [{\citenamefont {Diekmann}\ \emph {et~al.}(1995)\citenamefont
  {Diekmann}, \citenamefont {Metz},\ and\ \citenamefont
  {Heesterbeek}}]{Diekmann1995}%
  \BibitemOpen
  \bibfield  {author} {\bibinfo {author} {\bibfnamefont {O.}~\bibnamefont
  {Diekmann}}, \bibinfo {author} {\bibfnamefont {J.~A.~J.}\ \bibnamefont
  {Metz}}, \ and\ \bibinfo {author} {\bibfnamefont {J.~A.~P.}\ \bibnamefont
  {Heesterbeek}},\ }\bibfield  {title} {\enquote {\bibinfo {title} {{The legacy
  of Kermack and McKendrick}},}\ }in\ \href@noop {} {\emph {\bibinfo
  {booktitle} {Epidemic Model. Their Struct. Relat. to Data}}},\ \bibinfo
  {editor} {edited by\ \bibinfo {editor} {\bibfnamefont {D.}~\bibnamefont
  {Mollison}}}\ (\bibinfo  {publisher} {Cambridge University Press},\ \bibinfo
  {year} {1995})\ pp.\ \bibinfo {pages} {95--115}\BibitemShut {NoStop}%
\bibitem [{\citenamefont {Sheikh}\ \emph {et~al.}(2020)\citenamefont {Sheikh},
  \citenamefont {Watkins}, \citenamefont {Wu},\ and\ \citenamefont
  {Gr\"{o}ndahl}}]{nyt}%
  \BibitemOpen
  \bibfield  {author} {\bibinfo {author} {\bibfnamefont {K.}~\bibnamefont
  {Sheikh}}, \bibinfo {author} {\bibfnamefont {D.}~\bibnamefont {Watkins}},
  \bibinfo {author} {\bibfnamefont {J.}~\bibnamefont {Wu}}, \ and\ \bibinfo
  {author} {\bibfnamefont {M.}~\bibnamefont {Gr\"{o}ndahl}},\ }\bibfield
  {title} {\enquote {\bibinfo {title} {{How Bad Will the Coronavirus Outbreak
  Get? Here Are 6 Key Factors}},}\ }\href@noop {} {\bibfield  {journal}
  {\bibinfo  {journal} {The New York Times}\ } (\bibinfo {year}
  {2020})}\BibitemShut {NoStop}%
\bibitem [{\citenamefont {Kermack}\ and\ \citenamefont
  {McKendrick}(1927)}]{Kermack1927}%
  \BibitemOpen
  \bibfield  {author} {\bibinfo {author} {\bibfnamefont {W.~O.}\ \bibnamefont
  {Kermack}}\ and\ \bibinfo {author} {\bibfnamefont {A.~G.}\ \bibnamefont
  {McKendrick}},\ }\bibfield  {title} {\enquote {\bibinfo {title} {{A
  Contribution to the Mathematical Theory of Epidemics}},}\ }\href {\doibase
  10.1098/rspa.1927.0118} {\bibfield  {journal} {\bibinfo  {journal} {Proc. R.
  Soc. A}\ }\textbf {\bibinfo {volume} {115}},\ \bibinfo {pages} {700--721}
  (\bibinfo {year} {1927})}\BibitemShut {NoStop}%
\bibitem [{\citenamefont {Kermack}\ and\ \citenamefont
  {McKendrick}(1932)}]{Kermack1932}%
  \BibitemOpen
  \bibfield  {author} {\bibinfo {author} {\bibfnamefont {W.~O.}\ \bibnamefont
  {Kermack}}\ and\ \bibinfo {author} {\bibfnamefont {A.~G.}\ \bibnamefont
  {McKendrick}},\ }\bibfield  {title} {\enquote {\bibinfo {title}
  {{Contributions to the Mathematical Theory of Epidemics. II. The Problem of
  Endemicity}},}\ }\href {\doibase 10.1098/rspa.1932.0171} {\bibfield
  {journal} {\bibinfo  {journal} {Proc. R. Soc. A}\ }\textbf {\bibinfo {volume}
  {138}},\ \bibinfo {pages} {55--83} (\bibinfo {year} {1932})}\BibitemShut
  {NoStop}%
\bibitem [{\citenamefont {Kermack}\ and\ \citenamefont
  {McKendrick}(1933)}]{Kermack1933}%
  \BibitemOpen
  \bibfield  {author} {\bibinfo {author} {\bibfnamefont {W.~O.}\ \bibnamefont
  {Kermack}}\ and\ \bibinfo {author} {\bibfnamefont {A.~G.}\ \bibnamefont
  {McKendrick}},\ }\bibfield  {title} {\enquote {\bibinfo {title}
  {{Contributions to the Mathematical Theory of Epidemics. III. Further Studies
  of the Problem of Endemicity}},}\ }\href {\doibase 10.1098/rspa.1933.0106}
  {\bibfield  {journal} {\bibinfo  {journal} {Proc. R. Soc. A}\ }\textbf
  {\bibinfo {volume} {141}},\ \bibinfo {pages} {94--122} (\bibinfo {year}
  {1933})}\BibitemShut {NoStop}%
\bibitem [{\citenamefont {Meyers}\ \emph {et~al.}(2005)\citenamefont {Meyers},
  \citenamefont {Pourbohloul}, \citenamefont {Newman}, \citenamefont
  {Skowronski},\ and\ \citenamefont {Brunham}}]{Meyers2005}%
  \BibitemOpen
  \bibfield  {author} {\bibinfo {author} {\bibfnamefont {L.~A.}\ \bibnamefont
  {Meyers}}, \bibinfo {author} {\bibfnamefont {B.}~\bibnamefont {Pourbohloul}},
  \bibinfo {author} {\bibfnamefont {M.~E.~J.}\ \bibnamefont {Newman}}, \bibinfo
  {author} {\bibfnamefont {D.~M.}\ \bibnamefont {Skowronski}}, \ and\ \bibinfo
  {author} {\bibfnamefont {R.~C.}\ \bibnamefont {Brunham}},\ }\bibfield
  {title} {\enquote {\bibinfo {title} {{Network theory and SARS: Predicting
  outbreak diversity.}}}\ }\href {\doibase 10.1016/j.jtbi.2004.07.026}
  {\bibfield  {journal} {\bibinfo  {journal} {J. Theor. Biol.}\ }\textbf
  {\bibinfo {volume} {232}},\ \bibinfo {pages} {71--81} (\bibinfo {year}
  {2005})}\BibitemShut {NoStop}%
\bibitem [{\citenamefont {Memish}\ \emph {et~al.}(2014)\citenamefont {Memish},
  \citenamefont {Assiri}, \citenamefont {Almasri}, \citenamefont {Alhakeem},
  \citenamefont {Turkestani}, \citenamefont {{Al Rabeeah}}, \citenamefont
  {Al-Tawfiq}, \citenamefont {Alzahrani}, \citenamefont {Azhar}, \citenamefont
  {Makhdoom}, \citenamefont {Hajomar}, \citenamefont {Al-Shangiti},\ and\
  \citenamefont {Yezli}}]{Memish2014}%
  \BibitemOpen
  \bibfield  {author} {\bibinfo {author} {\bibfnamefont {Z.~A.}\ \bibnamefont
  {Memish}}, \bibinfo {author} {\bibfnamefont {A.}~\bibnamefont {Assiri}},
  \bibinfo {author} {\bibfnamefont {M.}~\bibnamefont {Almasri}}, \bibinfo
  {author} {\bibfnamefont {R.~F.}\ \bibnamefont {Alhakeem}}, \bibinfo {author}
  {\bibfnamefont {A.}~\bibnamefont {Turkestani}}, \bibinfo {author}
  {\bibfnamefont {A.~A.}\ \bibnamefont {{Al Rabeeah}}}, \bibinfo {author}
  {\bibfnamefont {J.~A.}\ \bibnamefont {Al-Tawfiq}}, \bibinfo {author}
  {\bibfnamefont {A.}~\bibnamefont {Alzahrani}}, \bibinfo {author}
  {\bibfnamefont {E.}~\bibnamefont {Azhar}}, \bibinfo {author} {\bibfnamefont
  {H.~Q.}\ \bibnamefont {Makhdoom}}, \bibinfo {author} {\bibfnamefont {W.~H.}\
  \bibnamefont {Hajomar}}, \bibinfo {author} {\bibfnamefont {A.~M.}\
  \bibnamefont {Al-Shangiti}}, \ and\ \bibinfo {author} {\bibfnamefont
  {S.}~\bibnamefont {Yezli}},\ }\bibfield  {title} {\enquote {\bibinfo {title}
  {{Prevalence of MERS-CoV Nasal Carriage and Compliance With the Saudi Health
  Recommendations Among Pilgrims Attending the 2013 Hajj}},}\ }\href {\doibase
  10.1093/infdis/jiu150} {\bibfield  {journal} {\bibinfo  {journal} {J. Infect.
  Dis.}\ }\textbf {\bibinfo {volume} {210}},\ \bibinfo {pages} {1067--1072}
  (\bibinfo {year} {2014})}\BibitemShut {NoStop}%
\bibitem [{\citenamefont {Kucharski}\ and\ \citenamefont
  {Althaus}(2015)}]{Kucharski2015}%
  \BibitemOpen
  \bibfield  {author} {\bibinfo {author} {\bibfnamefont {A.~J.}\ \bibnamefont
  {Kucharski}}\ and\ \bibinfo {author} {\bibfnamefont {C.~L.}\ \bibnamefont
  {Althaus}},\ }\bibfield  {title} {\enquote {\bibinfo {title} {{The role of
  superspreading in Middle East respiratory syndrome coronavirus (MERS-CoV)
  transmission}},}\ }\href {\doibase 10.2807/1560-7917.ES2015.20.25.21167}
  {\bibfield  {journal} {\bibinfo  {journal} {Eurosurveillance}\ }\textbf
  {\bibinfo {volume} {20}},\ \bibinfo {pages} {pii=21167} (\bibinfo {year}
  {2015})}\BibitemShut {NoStop}%
\bibitem [{\citenamefont {Leung}\ \emph {et~al.}(2006)\citenamefont {Leung},
  \citenamefont {Lim}, \citenamefont {Ho}, \citenamefont {Lam}, \citenamefont
  {Ghani}, \citenamefont {Donnelly}, \citenamefont {Fraser}, \citenamefont
  {Riley}, \citenamefont {Ferguson}, \citenamefont {Anderson},\ and\
  \citenamefont {Hedley}}]{leung_seroprevalence_2006}%
  \BibitemOpen
  \bibfield  {author} {\bibinfo {author} {\bibfnamefont {G.~M.}\ \bibnamefont
  {Leung}}, \bibinfo {author} {\bibfnamefont {W.~W.}\ \bibnamefont {Lim}},
  \bibinfo {author} {\bibfnamefont {L.-M.}\ \bibnamefont {Ho}}, \bibinfo
  {author} {\bibfnamefont {T.-H.}\ \bibnamefont {Lam}}, \bibinfo {author}
  {\bibfnamefont {A.~C.}\ \bibnamefont {Ghani}}, \bibinfo {author}
  {\bibfnamefont {C.~A.}\ \bibnamefont {Donnelly}}, \bibinfo {author}
  {\bibfnamefont {C.}~\bibnamefont {Fraser}}, \bibinfo {author} {\bibfnamefont
  {S.}~\bibnamefont {Riley}}, \bibinfo {author} {\bibfnamefont {N.~M.}\
  \bibnamefont {Ferguson}}, \bibinfo {author} {\bibfnamefont {R.~M.}\
  \bibnamefont {Anderson}}, \ and\ \bibinfo {author} {\bibfnamefont {A.~J.}\
  \bibnamefont {Hedley}},\ }\bibfield  {title} {\enquote {\bibinfo {title}
  {Seroprevalence of {IgG} antibodies to {SARS}-coronavirus in asymptomatic or
  subclinical population groups},}\ }\href {\doibase 10.1017/S0950268805004826}
  {\bibfield  {journal} {\bibinfo  {journal} {Epidemiology and Infection}\
  }\textbf {\bibinfo {volume} {134}},\ \bibinfo {pages} {211--221} (\bibinfo
  {year} {2006})}\BibitemShut {NoStop}%
\bibitem [{\citenamefont {Quah}\ and\ \citenamefont
  {Hin-Peng}(2004)}]{Quah2004}%
  \BibitemOpen
  \bibfield  {author} {\bibinfo {author} {\bibfnamefont {S.~R.}\ \bibnamefont
  {Quah}}\ and\ \bibinfo {author} {\bibfnamefont {L.}~\bibnamefont
  {Hin-Peng}},\ }\bibfield  {title} {\enquote {\bibinfo {title} {{Crisis
  Prevention and Management during SARS Outbreak, Singapore}},}\ }\href
  {\doibase 10.3201/eid1002.030418} {\bibfield  {journal} {\bibinfo  {journal}
  {Emerg. Infect. Dis.}\ }\textbf {\bibinfo {volume} {10}},\ \bibinfo {pages}
  {364--368} (\bibinfo {year} {2004})}\BibitemShut {NoStop}%
\bibitem [{\citenamefont {Mack}\ \emph {et~al.}(1972)\citenamefont {Mack},
  \citenamefont {Thoma}, \citenamefont {Ali},\ and\ \citenamefont
  {Khan}}]{Mack1972}%
  \BibitemOpen
  \bibfield  {author} {\bibinfo {author} {\bibfnamefont {T.~M.}\ \bibnamefont
  {Mack}}, \bibinfo {author} {\bibfnamefont {D.~B.}\ \bibnamefont {Thoma}},
  \bibinfo {author} {\bibfnamefont {A.}~\bibnamefont {Ali}}, \ and\ \bibinfo
  {author} {\bibfnamefont {M.~M.}\ \bibnamefont {Khan}},\ }\bibfield  {title}
  {\enquote {\bibinfo {title} {{Epidemiology of smallpox in West Pakistan}},}\
  }\href {\doibase 10.1093/oxfordjournals.aje.a121380} {\bibfield  {journal}
  {\bibinfo  {journal} {Am. J. Epidemiol.}\ }\textbf {\bibinfo {volume} {95}},\
  \bibinfo {pages} {157--168} (\bibinfo {year} {1972})}\BibitemShut {NoStop}%
\bibitem [{\citenamefont {Taubenberger}\ and\ \citenamefont
  {Morens}(2006)}]{Taubenberger2006}%
  \BibitemOpen
  \bibfield  {author} {\bibinfo {author} {\bibfnamefont {J.~K.}\ \bibnamefont
  {Taubenberger}}\ and\ \bibinfo {author} {\bibfnamefont {D.~M.}\ \bibnamefont
  {Morens}},\ }\bibfield  {title} {\enquote {\bibinfo {title} {{1918 Influenza:
  the Mother of All Pandemics}},}\ }\href {\doibase 10.3201/eid1201.050979}
  {\bibfield  {journal} {\bibinfo  {journal} {Emerg. Infect. Dis.}\ }\textbf
  {\bibinfo {volume} {12}},\ \bibinfo {pages} {15--22} (\bibinfo {year}
  {2006})}\BibitemShut {NoStop}%
\bibitem [{\citenamefont {Fraser}\ \emph {et~al.}(2011)\citenamefont {Fraser},
  \citenamefont {Cummings}, \citenamefont {Klinkenberg}, \citenamefont
  {Burke},\ and\ \citenamefont {Ferguson}}]{Fraser2011}%
  \BibitemOpen
  \bibfield  {author} {\bibinfo {author} {\bibfnamefont {C.}~\bibnamefont
  {Fraser}}, \bibinfo {author} {\bibfnamefont {D.~A.~T.}\ \bibnamefont
  {Cummings}}, \bibinfo {author} {\bibfnamefont {D.}~\bibnamefont
  {Klinkenberg}}, \bibinfo {author} {\bibfnamefont {D.~S.}\ \bibnamefont
  {Burke}}, \ and\ \bibinfo {author} {\bibfnamefont {N.~M.}\ \bibnamefont
  {Ferguson}},\ }\bibfield  {title} {\enquote {\bibinfo {title} {{Influenza
  Transmission in Households During the 1918 Pandemic}},}\ }\href {\doibase
  10.1093/aje/kwr122} {\bibfield  {journal} {\bibinfo  {journal} {Am. J.
  Epidemiol.}\ }\textbf {\bibinfo {volume} {174}},\ \bibinfo {pages} {505--514}
  (\bibinfo {year} {2011})}\BibitemShut {NoStop}%
\bibitem [{\citenamefont {Kelly}\ \emph {et~al.}(2011)\citenamefont {Kelly},
  \citenamefont {Peck}, \citenamefont {Laurie}, \citenamefont {Wu},
  \citenamefont {Nishiura},\ and\ \citenamefont {Cowling}}]{Kelly2011}%
  \BibitemOpen
  \bibfield  {author} {\bibinfo {author} {\bibfnamefont {H.}~\bibnamefont
  {Kelly}}, \bibinfo {author} {\bibfnamefont {H.~A.}\ \bibnamefont {Peck}},
  \bibinfo {author} {\bibfnamefont {K.~L.}\ \bibnamefont {Laurie}}, \bibinfo
  {author} {\bibfnamefont {P.}~\bibnamefont {Wu}}, \bibinfo {author}
  {\bibfnamefont {H.}~\bibnamefont {Nishiura}}, \ and\ \bibinfo {author}
  {\bibfnamefont {B.~J.}\ \bibnamefont {Cowling}},\ }\bibfield  {title}
  {\enquote {\bibinfo {title} {{The Age-Specific Cumulative Incidence of
  Infection with Pandemic Influenza H1N1 2009 Was Similar in Various Countries
  Prior to Vaccination}},}\ }\href {\doibase 10.1371/journal.pone.0021828}
  {\bibfield  {journal} {\bibinfo  {journal} {PLOS ONE}\ }\textbf {\bibinfo
  {volume} {6}},\ \bibinfo {pages} {e21828} (\bibinfo {year}
  {2011})}\BibitemShut {NoStop}%
\bibitem [{\citenamefont {Dorigatti}\ \emph {et~al.}(2012)\citenamefont
  {Dorigatti}, \citenamefont {Cauchemez}, \citenamefont {Pugliese},\ and\
  \citenamefont {Ferguson}}]{Dorigatti2012}%
  \BibitemOpen
  \bibfield  {author} {\bibinfo {author} {\bibfnamefont {I.}~\bibnamefont
  {Dorigatti}}, \bibinfo {author} {\bibfnamefont {S.}~\bibnamefont
  {Cauchemez}}, \bibinfo {author} {\bibfnamefont {A.}~\bibnamefont {Pugliese}},
  \ and\ \bibinfo {author} {\bibfnamefont {N.~M.}\ \bibnamefont {Ferguson}},\
  }\bibfield  {title} {\enquote {\bibinfo {title} {{A new approach to
  characterising infectious disease transmission dynamics from sentinel
  surveillance: Application to the Italian 2009–2010 A/H1N1 influenza
  pandemic}},}\ }\href {\doibase 10.1016/j.epidem.2011.11.001} {\bibfield
  {journal} {\bibinfo  {journal} {Epidemics}\ }\textbf {\bibinfo {volume}
  {4}},\ \bibinfo {pages} {9--21} (\bibinfo {year} {2012})}\BibitemShut
  {NoStop}%
\bibitem [{\citenamefont {Li}\ \emph {et~al.}(2020)\citenamefont {Li},
  \citenamefont {Guan}, \citenamefont {Wu}, \citenamefont {Wang}, \citenamefont
  {Zhou}, \citenamefont {Tong}, \citenamefont {Ren}, \citenamefont {Leung},
  \citenamefont {Lau}, \citenamefont {Wong}, \citenamefont {Xing},
  \citenamefont {Xiang}, \citenamefont {Wu}, \citenamefont {Li}, \citenamefont
  {Chen}, \citenamefont {Li}, \citenamefont {Liu}, \citenamefont {Zhao},
  \citenamefont {Liu}, \citenamefont {Tu}, \citenamefont {Chen}, \citenamefont
  {Jin}, \citenamefont {Yang}, \citenamefont {Wang}, \citenamefont {Zhou},
  \citenamefont {Wang}, \citenamefont {Liu}, \citenamefont {Luo}, \citenamefont
  {Liu}, \citenamefont {Shao}, \citenamefont {Li}, \citenamefont {Tao},
  \citenamefont {Yang}, \citenamefont {Deng}, \citenamefont {Liu},
  \citenamefont {Ma}, \citenamefont {Zhang}, \citenamefont {Shi}, \citenamefont
  {Lam}, \citenamefont {Wu}, \citenamefont {Gao}, \citenamefont {Cowling},
  \citenamefont {Yang}, \citenamefont {Leung},\ and\ \citenamefont
  {Feng}}]{Li2020}%
  \BibitemOpen
  \bibfield  {author} {\bibinfo {author} {\bibfnamefont {Q.}~\bibnamefont
  {Li}}, \bibinfo {author} {\bibfnamefont {X.}~\bibnamefont {Guan}}, \bibinfo
  {author} {\bibfnamefont {P.}~\bibnamefont {Wu}}, \bibinfo {author}
  {\bibfnamefont {X.}~\bibnamefont {Wang}}, \bibinfo {author} {\bibfnamefont
  {L.}~\bibnamefont {Zhou}}, \bibinfo {author} {\bibfnamefont {Y.}~\bibnamefont
  {Tong}}, \bibinfo {author} {\bibfnamefont {R.}~\bibnamefont {Ren}}, \bibinfo
  {author} {\bibfnamefont {K.~S.~M.}\ \bibnamefont {Leung}}, \bibinfo {author}
  {\bibfnamefont {E.~H.~Y.}\ \bibnamefont {Lau}}, \bibinfo {author}
  {\bibfnamefont {J.~Y.}\ \bibnamefont {Wong}}, \bibinfo {author}
  {\bibfnamefont {X.}~\bibnamefont {Xing}}, \bibinfo {author} {\bibfnamefont
  {N.}~\bibnamefont {Xiang}}, \bibinfo {author} {\bibfnamefont
  {Y.}~\bibnamefont {Wu}}, \bibinfo {author} {\bibfnamefont {C.}~\bibnamefont
  {Li}}, \bibinfo {author} {\bibfnamefont {Q.}~\bibnamefont {Chen}}, \bibinfo
  {author} {\bibfnamefont {D.}~\bibnamefont {Li}}, \bibinfo {author}
  {\bibfnamefont {T.}~\bibnamefont {Liu}}, \bibinfo {author} {\bibfnamefont
  {J.}~\bibnamefont {Zhao}}, \bibinfo {author} {\bibfnamefont {M.}~\bibnamefont
  {Liu}}, \bibinfo {author} {\bibfnamefont {W.}~\bibnamefont {Tu}}, \bibinfo
  {author} {\bibfnamefont {C.}~\bibnamefont {Chen}}, \bibinfo {author}
  {\bibfnamefont {L.}~\bibnamefont {Jin}}, \bibinfo {author} {\bibfnamefont
  {R.}~\bibnamefont {Yang}}, \bibinfo {author} {\bibfnamefont {Q.}~\bibnamefont
  {Wang}}, \bibinfo {author} {\bibfnamefont {S.}~\bibnamefont {Zhou}}, \bibinfo
  {author} {\bibfnamefont {R.}~\bibnamefont {Wang}}, \bibinfo {author}
  {\bibfnamefont {H.}~\bibnamefont {Liu}}, \bibinfo {author} {\bibfnamefont
  {Y.}~\bibnamefont {Luo}}, \bibinfo {author} {\bibfnamefont {Y.}~\bibnamefont
  {Liu}}, \bibinfo {author} {\bibfnamefont {G.}~\bibnamefont {Shao}}, \bibinfo
  {author} {\bibfnamefont {H.}~\bibnamefont {Li}}, \bibinfo {author}
  {\bibfnamefont {Z.}~\bibnamefont {Tao}}, \bibinfo {author} {\bibfnamefont
  {Y.}~\bibnamefont {Yang}}, \bibinfo {author} {\bibfnamefont {Z.}~\bibnamefont
  {Deng}}, \bibinfo {author} {\bibfnamefont {B.}~\bibnamefont {Liu}}, \bibinfo
  {author} {\bibfnamefont {Z.}~\bibnamefont {Ma}}, \bibinfo {author}
  {\bibfnamefont {Y.}~\bibnamefont {Zhang}}, \bibinfo {author} {\bibfnamefont
  {G.}~\bibnamefont {Shi}}, \bibinfo {author} {\bibfnamefont {T.~T.~Y.}\
  \bibnamefont {Lam}}, \bibinfo {author} {\bibfnamefont {J.~T.}\ \bibnamefont
  {Wu}}, \bibinfo {author} {\bibfnamefont {G.~F.}\ \bibnamefont {Gao}},
  \bibinfo {author} {\bibfnamefont {B.~J.}\ \bibnamefont {Cowling}}, \bibinfo
  {author} {\bibfnamefont {B.}~\bibnamefont {Yang}}, \bibinfo {author}
  {\bibfnamefont {G.~M.}\ \bibnamefont {Leung}}, \ and\ \bibinfo {author}
  {\bibfnamefont {Z.}~\bibnamefont {Feng}},\ }\bibfield  {title} {\enquote
  {\bibinfo {title} {{Early Transmission Dynamics in Wuhan, China, of Novel
  Coronavirus–Infected Pneumonia}},}\ }\href {\doibase 10.1056/NEJMoa2001316}
  {\bibfield  {journal} {\bibinfo  {journal} {N. Engl. J. Med.}\ } (\bibinfo
  {year} {2020}),\ 10.1056/NEJMoa2001316}\BibitemShut {NoStop}%
\bibitem [{\citenamefont {Endo}\ \emph {et~al.}(2020)\citenamefont {Endo},
  \citenamefont {{Centre for the Mathematical Modelling of Infectious Diseases
  COVID-19 Working Group}}, \citenamefont {Abbott}, \citenamefont {Kucharski},\
  and\ \citenamefont {Funk}}]{endo_estimating_2020}%
  \BibitemOpen
  \bibfield  {author} {\bibinfo {author} {\bibfnamefont {A.}~\bibnamefont
  {Endo}}, \bibinfo {author} {\bibnamefont {{Centre for the Mathematical
  Modelling of Infectious Diseases COVID-19 Working Group}}}, \bibinfo {author}
  {\bibfnamefont {S.}~\bibnamefont {Abbott}}, \bibinfo {author} {\bibfnamefont
  {A.~J.}\ \bibnamefont {Kucharski}}, \ and\ \bibinfo {author} {\bibfnamefont
  {S.}~\bibnamefont {Funk}},\ }\bibfield  {title} {\enquote {\bibinfo {title}
  {{Estimating the overdispersion in {COVID}-19 transmission using outbreak
  sizes outside {\{}China{\}}}},}\ }\href {\doibase
  10.12688/wellcomeopenres.15842.1} {\bibfield  {journal} {\bibinfo  {journal}
  {Wellcome Open Res.}\ }\textbf {\bibinfo {volume} {5}},\ \bibinfo {pages}
  {67} (\bibinfo {year} {2020})}\BibitemShut {NoStop}%
\bibitem [{\citenamefont {Streeck}\ \emph {et~al.}(2020)\citenamefont
  {Streeck}, \citenamefont {Hartmann}, \citenamefont {Exner},\ and\
  \citenamefont {Schmid}}]{Streeck2020}%
  \BibitemOpen
  \bibfield  {author} {\bibinfo {author} {\bibfnamefont {H.}~\bibnamefont
  {Streeck}}, \bibinfo {author} {\bibfnamefont {G.}~\bibnamefont {Hartmann}},
  \bibinfo {author} {\bibfnamefont {M.}~\bibnamefont {Exner}}, \ and\ \bibinfo
  {author} {\bibfnamefont {M.}~\bibnamefont {Schmid}},\ }\bibfield  {title}
  {\enquote {\bibinfo {title} {Preliminary result and conclusions of the
  {COVID}-19 case cluster study ({G}angelt {M}unicipality)},}\ }\href
  {https://www.land.nrw/sites/default/files/asset/document/zwischenergebnis_covid19_case_study_gangelt_en.pdf}
  {\bibfield  {journal} {\bibinfo  {journal} {{online preprint}}\ } (\bibinfo
  {year} {2020})}\BibitemShut {NoStop}%
\bibitem [{\citenamefont {Sutton}\ \emph {et~al.}(2020)\citenamefont {Sutton},
  \citenamefont {Fuchs}, \citenamefont {D’Alton},\ and\ \citenamefont
  {Goffman}}]{sutton_universal_2020}%
  \BibitemOpen
  \bibfield  {author} {\bibinfo {author} {\bibfnamefont {D.}~\bibnamefont
  {Sutton}}, \bibinfo {author} {\bibfnamefont {K.}~\bibnamefont {Fuchs}},
  \bibinfo {author} {\bibfnamefont {M.}~\bibnamefont {D’Alton}}, \ and\
  \bibinfo {author} {\bibfnamefont {D.}~\bibnamefont {Goffman}},\ }\bibfield
  {title} {\enquote {\bibinfo {title} {Universal {Screening} for {SARS}-{CoV}-2
  in {Women} {Admitted} for {Delivery}},}\ }\href {\doibase
  10.1056/NEJMc2009316} {\bibfield  {journal} {\bibinfo  {journal} {New England
  Journal of Medicine}\ ,\ \bibinfo {pages} {NEJMc2009316}} (\bibinfo {year}
  {2020})}\BibitemShut {NoStop}%
\bibitem [{\citenamefont {Eksin}\ \emph {et~al.}(2019)\citenamefont {Eksin},
  \citenamefont {Paarporn},\ and\ \citenamefont {Weitz}}]{Eksin2019}%
  \BibitemOpen
  \bibfield  {author} {\bibinfo {author} {\bibfnamefont {C.}~\bibnamefont
  {Eksin}}, \bibinfo {author} {\bibfnamefont {K.}~\bibnamefont {Paarporn}}, \
  and\ \bibinfo {author} {\bibfnamefont {J.~S.}\ \bibnamefont {Weitz}},\
  }\bibfield  {title} {\enquote {\bibinfo {title} {{Systematic biases in
  disease forecasting – The role of behavior change}},}\ }\href {\doibase
  10.1016/j.epidem.2019.02.004} {\bibfield  {journal} {\bibinfo  {journal}
  {Epidemics}\ }\textbf {\bibinfo {volume} {27}},\ \bibinfo {pages} {96--105}
  (\bibinfo {year} {2019})}\BibitemShut {NoStop}%
\bibitem [{WHO(2019)}]{WHO19}%
  \BibitemOpen
  \href@noop {} {\enquote {\bibinfo {title} {{WHO} disease outbreaks by year:
  2019},}\ }\bibinfo {howpublished}
  {\url{https://www.who.int/csr/don/archive/year/2019/en/}} (\bibinfo {year}
  {2019}),\ \bibinfo {note} {accessed: 2020-02-09}\BibitemShut {NoStop}%
\bibitem [{\citenamefont {Moreno}\ \emph {et~al.}(2002)\citenamefont {Moreno},
  \citenamefont {Pastor-Satorras},\ and\ \citenamefont
  {Vespignani}}]{Moreno2002}%
  \BibitemOpen
  \bibfield  {author} {\bibinfo {author} {\bibfnamefont {Y.}~\bibnamefont
  {Moreno}}, \bibinfo {author} {\bibfnamefont {R.}~\bibnamefont
  {Pastor-Satorras}}, \ and\ \bibinfo {author} {\bibfnamefont {A.}~\bibnamefont
  {Vespignani}},\ }\bibfield  {title} {\enquote {\bibinfo {title} {{Epidemic
  outbreaks in complex heterogeneous networks}},}\ }\href {\doibase
  10.1140/epjb/e20020122} {\bibfield  {journal} {\bibinfo  {journal} {Eur.
  Phys. J. B}\ }\textbf {\bibinfo {volume} {26}},\ \bibinfo {pages} {521--529}
  (\bibinfo {year} {2002})}\BibitemShut {NoStop}%
\bibitem [{\citenamefont {Colizza}\ and\ \citenamefont
  {Vespignani}(2008)}]{Colizza2008}%
  \BibitemOpen
  \bibfield  {author} {\bibinfo {author} {\bibfnamefont {V.}~\bibnamefont
  {Colizza}}\ and\ \bibinfo {author} {\bibfnamefont {A.}~\bibnamefont
  {Vespignani}},\ }\bibfield  {title} {\enquote {\bibinfo {title} {{Epidemic
  modeling in metapopulation systems with heterogeneous coupling pattern:
  Theory and simulations}},}\ }\href {\doibase 10.1016/j.jtbi.2007.11.028}
  {\bibfield  {journal} {\bibinfo  {journal} {J. Theor. Biol.}\ }\textbf
  {\bibinfo {volume} {251}},\ \bibinfo {pages} {450--467} (\bibinfo {year}
  {2008})}\BibitemShut {NoStop}%
\bibitem [{\citenamefont {Wesolowski}\ \emph {et~al.}(2015)\citenamefont
  {Wesolowski}, \citenamefont {Qureshi}, \citenamefont {Boni}, \citenamefont
  {Sunds{\o}y}, \citenamefont {Johansson}, \citenamefont {Rasheed},
  \citenamefont {Eng{\o}-Monsen},\ and\ \citenamefont
  {Buckee}}]{Wesolowski2015}%
  \BibitemOpen
  \bibfield  {author} {\bibinfo {author} {\bibfnamefont {A.}~\bibnamefont
  {Wesolowski}}, \bibinfo {author} {\bibfnamefont {T.}~\bibnamefont {Qureshi}},
  \bibinfo {author} {\bibfnamefont {M.~F.}\ \bibnamefont {Boni}}, \bibinfo
  {author} {\bibfnamefont {P.~R.}\ \bibnamefont {Sunds{\o}y}}, \bibinfo
  {author} {\bibfnamefont {M.~A.}\ \bibnamefont {Johansson}}, \bibinfo {author}
  {\bibfnamefont {S.~B.}\ \bibnamefont {Rasheed}}, \bibinfo {author}
  {\bibfnamefont {K.}~\bibnamefont {Eng{\o}-Monsen}}, \ and\ \bibinfo {author}
  {\bibfnamefont {C.~O.}\ \bibnamefont {Buckee}},\ }\bibfield  {title}
  {\enquote {\bibinfo {title} {{Impact of human mobility on the emergence of
  dengue epidemics in Pakistan}},}\ }\href {\doibase 10.1073/pnas.1504964112}
  {\bibfield  {journal} {\bibinfo  {journal} {Proc. Natl. Acad. Sci. USA}\
  }\textbf {\bibinfo {volume} {112}},\ \bibinfo {pages} {11887--11892}
  (\bibinfo {year} {2015})}\BibitemShut {NoStop}%
\bibitem [{\citenamefont {Scarpino}\ \emph {et~al.}(2016)\citenamefont
  {Scarpino}, \citenamefont {Allard},\ and\ \citenamefont
  {H{\'{e}}bert-Dufresne}}]{Scarpino2016}%
  \BibitemOpen
  \bibfield  {author} {\bibinfo {author} {\bibfnamefont {S.~V.}\ \bibnamefont
  {Scarpino}}, \bibinfo {author} {\bibfnamefont {A.}~\bibnamefont {Allard}}, \
  and\ \bibinfo {author} {\bibfnamefont {L.}~\bibnamefont
  {H{\'{e}}bert-Dufresne}},\ }\bibfield  {title} {\enquote {\bibinfo {title}
  {{The effect of a prudent adaptive behaviour on disease transmission}},}\
  }\href {\doibase 10.1038/nphys3832} {\bibfield  {journal} {\bibinfo
  {journal} {Nat. Phys.}\ }\textbf {\bibinfo {volume} {12}},\ \bibinfo {pages}
  {1042--1046} (\bibinfo {year} {2016})}\BibitemShut {NoStop}%
\bibitem [{\citenamefont {H{\'{e}}bert-Dufresne}\ and\ \citenamefont
  {Althouse}(2015)}]{Hebert-Dufresne2015}%
  \BibitemOpen
  \bibfield  {author} {\bibinfo {author} {\bibfnamefont {L.}~\bibnamefont
  {H{\'{e}}bert-Dufresne}}\ and\ \bibinfo {author} {\bibfnamefont {B.~M.}\
  \bibnamefont {Althouse}},\ }\bibfield  {title} {\enquote {\bibinfo {title}
  {{Complex dynamics of synergistic coinfections on realistically clustered
  networks}},}\ }\href {\doibase 10.1073/pnas.1507820112} {\bibfield  {journal}
  {\bibinfo  {journal} {Proc. Natl. Acad. Sci. USA}\ }\textbf {\bibinfo
  {volume} {112}},\ \bibinfo {pages} {10551--10556} (\bibinfo {year}
  {2015})}\BibitemShut {NoStop}%
\bibitem [{\citenamefont {H{\'{e}}bert-Dufresne}\ \emph
  {et~al.}(2019)\citenamefont {H{\'{e}}bert-Dufresne}, \citenamefont
  {Scarpino},\ and\ \citenamefont {Young}}]{Hebert-Dufresne2019}%
  \BibitemOpen
  \bibfield  {author} {\bibinfo {author} {\bibfnamefont {L.}~\bibnamefont
  {H{\'{e}}bert-Dufresne}}, \bibinfo {author} {\bibfnamefont {S.~V.}\
  \bibnamefont {Scarpino}}, \ and\ \bibinfo {author} {\bibfnamefont {J.-G.}\
  \bibnamefont {Young}},\ }\bibfield  {title} {\enquote {\bibinfo {title}
  {{Interacting contagions are indistinguishable from social reinforcement}},}\
  }\href@noop {} {\bibfield  {journal} {\bibinfo  {journal} {arXiv}\ }
  (\bibinfo {year} {2019})},\ \Eprint {http://arxiv.org/abs/1906.01147}
  {arXiv:1906.01147} \BibitemShut {NoStop}%
\bibitem [{\citenamefont {Dhillon}\ and\ \citenamefont
  {Srikrishna}(2018)}]{Dhillon2018}%
  \BibitemOpen
  \bibfield  {author} {\bibinfo {author} {\bibfnamefont {R.~S.}\ \bibnamefont
  {Dhillon}}\ and\ \bibinfo {author} {\bibfnamefont {D.}~\bibnamefont
  {Srikrishna}},\ }\bibfield  {title} {\enquote {\bibinfo {title} {{When is
  contact tracing not enough to stop an outbreak?}}}\ }\href {\doibase
  10.1016/S1473-3099(18)30656-X} {\bibfield  {journal} {\bibinfo  {journal}
  {Lancet Infect. Dis.}\ }\textbf {\bibinfo {volume} {18}},\ \bibinfo {pages}
  {1302--1304} (\bibinfo {year} {2018})}\BibitemShut {NoStop}%
\bibitem [{\citenamefont {Klinkenberg}\ \emph {et~al.}(2006)\citenamefont
  {Klinkenberg}, \citenamefont {Fraser},\ and\ \citenamefont
  {Heesterbeek}}]{Klinkenberg2006}%
  \BibitemOpen
  \bibfield  {author} {\bibinfo {author} {\bibfnamefont {D.}~\bibnamefont
  {Klinkenberg}}, \bibinfo {author} {\bibfnamefont {C.}~\bibnamefont {Fraser}},
  \ and\ \bibinfo {author} {\bibfnamefont {H.}~\bibnamefont {Heesterbeek}},\
  }\bibfield  {title} {\enquote {\bibinfo {title} {{The Effectiveness of
  Contact Tracing in Emerging Epidemics}},}\ }\href {\doibase
  10.1371/journal.pone.0000012} {\bibfield  {journal} {\bibinfo  {journal}
  {PLOS ONE}\ }\textbf {\bibinfo {volume} {1}},\ \bibinfo {pages} {e12}
  (\bibinfo {year} {2006})}\BibitemShut {NoStop}%
\bibitem [{\citenamefont {Smith}\ \emph {et~al.}(2009)\citenamefont {Smith},
  \citenamefont {Vijaykrishna}, \citenamefont {Bahl}, \citenamefont {Lycett},
  \citenamefont {Worobey}, \citenamefont {Pybus}, \citenamefont {Ma},
  \citenamefont {Cheung}, \citenamefont {Raghwani}, \citenamefont {Bhatt},
  \citenamefont {Peiris}, \citenamefont {Guan},\ and\ \citenamefont
  {Rambaut}}]{Smith2009}%
  \BibitemOpen
  \bibfield  {author} {\bibinfo {author} {\bibfnamefont {G.~J.~D.}\
  \bibnamefont {Smith}}, \bibinfo {author} {\bibfnamefont {D.}~\bibnamefont
  {Vijaykrishna}}, \bibinfo {author} {\bibfnamefont {J.}~\bibnamefont {Bahl}},
  \bibinfo {author} {\bibfnamefont {S.~J.}\ \bibnamefont {Lycett}}, \bibinfo
  {author} {\bibfnamefont {M.}~\bibnamefont {Worobey}}, \bibinfo {author}
  {\bibfnamefont {O.~G.}\ \bibnamefont {Pybus}}, \bibinfo {author}
  {\bibfnamefont {S.~K.}\ \bibnamefont {Ma}}, \bibinfo {author} {\bibfnamefont
  {C.~L.}\ \bibnamefont {Cheung}}, \bibinfo {author} {\bibfnamefont
  {J.}~\bibnamefont {Raghwani}}, \bibinfo {author} {\bibfnamefont
  {S.}~\bibnamefont {Bhatt}}, \bibinfo {author} {\bibfnamefont {J.~S.~M.}\
  \bibnamefont {Peiris}}, \bibinfo {author} {\bibfnamefont {Y.}~\bibnamefont
  {Guan}}, \ and\ \bibinfo {author} {\bibfnamefont {A.}~\bibnamefont
  {Rambaut}},\ }\bibfield  {title} {\enquote {\bibinfo {title} {{Origins and
  evolutionary genomics of the 2009 swine-origin H1N1 influenza A epidemic}},}\
  }\href {\doibase 10.1038/nature08182} {\bibfield  {journal} {\bibinfo
  {journal} {Nature}\ }\textbf {\bibinfo {volume} {459}},\ \bibinfo {pages}
  {1122--1125} (\bibinfo {year} {2009})}\BibitemShut {NoStop}%
\bibitem [{\citenamefont {Scarpino}\ \emph {et~al.}(2015)\citenamefont
  {Scarpino}, \citenamefont {Iamarino}, \citenamefont {Wells}, \citenamefont
  {Yamin}, \citenamefont {Ndeffo-Mbah}, \citenamefont {Wenzel}, \citenamefont
  {Fox}, \citenamefont {Nyenswah}, \citenamefont {Altice}, \citenamefont
  {Galvani}, \citenamefont {Meyers},\ and\ \citenamefont
  {Townsend}}]{Scarpino2015}%
  \BibitemOpen
  \bibfield  {author} {\bibinfo {author} {\bibfnamefont {S.~V.}\ \bibnamefont
  {Scarpino}}, \bibinfo {author} {\bibfnamefont {A.}~\bibnamefont {Iamarino}},
  \bibinfo {author} {\bibfnamefont {C.}~\bibnamefont {Wells}}, \bibinfo
  {author} {\bibfnamefont {D.}~\bibnamefont {Yamin}}, \bibinfo {author}
  {\bibfnamefont {M.}~\bibnamefont {Ndeffo-Mbah}}, \bibinfo {author}
  {\bibfnamefont {N.~S.}\ \bibnamefont {Wenzel}}, \bibinfo {author}
  {\bibfnamefont {S.~J.}\ \bibnamefont {Fox}}, \bibinfo {author} {\bibfnamefont
  {T.}~\bibnamefont {Nyenswah}}, \bibinfo {author} {\bibfnamefont {F.~L.}\
  \bibnamefont {Altice}}, \bibinfo {author} {\bibfnamefont {A.~P.}\
  \bibnamefont {Galvani}}, \bibinfo {author} {\bibfnamefont {L.~A.}\
  \bibnamefont {Meyers}}, \ and\ \bibinfo {author} {\bibfnamefont {Jeffrey~P.}\
  \bibnamefont {Townsend}},\ }\bibfield  {title} {\enquote {\bibinfo {title}
  {{Epidemiological and Viral Genomic Sequence Analysis of the 2014 Ebola
  Outbreak Reveals Clustered Transmission}},}\ }\href {\doibase
  10.1093/cid/ciu1131} {\bibfield  {journal} {\bibinfo  {journal} {Clin.
  Infect. Dis.}\ }\textbf {\bibinfo {volume} {60}},\ \bibinfo {pages}
  {1079--1082} (\bibinfo {year} {2015})}\BibitemShut {NoStop}%
\bibitem [{\citenamefont {Jombart}\ \emph {et~al.}(2014)\citenamefont
  {Jombart}, \citenamefont {Cori}, \citenamefont {Didelot}, \citenamefont
  {Cauchemez}, \citenamefont {Fraser},\ and\ \citenamefont
  {Ferguson}}]{Jombart2014}%
  \BibitemOpen
  \bibfield  {author} {\bibinfo {author} {\bibfnamefont {T.}~\bibnamefont
  {Jombart}}, \bibinfo {author} {\bibfnamefont {A.}~\bibnamefont {Cori}},
  \bibinfo {author} {\bibfnamefont {X.}~\bibnamefont {Didelot}}, \bibinfo
  {author} {\bibfnamefont {S.}~\bibnamefont {Cauchemez}}, \bibinfo {author}
  {\bibfnamefont {C.}~\bibnamefont {Fraser}}, \ and\ \bibinfo {author}
  {\bibfnamefont {N.}~\bibnamefont {Ferguson}},\ }\bibfield  {title} {\enquote
  {\bibinfo {title} {{Bayesian Reconstruction of Disease Outbreaks by Combining
  Epidemiologic and Genomic Data}},}\ }\href {\doibase
  10.1371/journal.pcbi.1003457} {\bibfield  {journal} {\bibinfo  {journal}
  {PLOS Comput. Biol.}\ }\textbf {\bibinfo {volume} {10}},\ \bibinfo {pages}
  {e1003457} (\bibinfo {year} {2014})}\BibitemShut {NoStop}%
\bibitem [{\citenamefont {Campbell}\ \emph {et~al.}(2019)\citenamefont
  {Campbell}, \citenamefont {Cori}, \citenamefont {Ferguson},\ and\
  \citenamefont {Jombart}}]{Campbell2019}%
  \BibitemOpen
  \bibfield  {author} {\bibinfo {author} {\bibfnamefont {F.}~\bibnamefont
  {Campbell}}, \bibinfo {author} {\bibfnamefont {A.}~\bibnamefont {Cori}},
  \bibinfo {author} {\bibfnamefont {N.}~\bibnamefont {Ferguson}}, \ and\
  \bibinfo {author} {\bibfnamefont {T.}~\bibnamefont {Jombart}},\ }\bibfield
  {title} {\enquote {\bibinfo {title} {{Bayesian inference of transmission
  chains using timing of symptoms, pathogen genomes and contact data}},}\
  }\href {\doibase 10.1371/journal.pcbi.1006930} {\bibfield  {journal}
  {\bibinfo  {journal} {PLOS Comput. Biol.}\ }\textbf {\bibinfo {volume}
  {15}},\ \bibinfo {pages} {e1006930} (\bibinfo {year} {2019})}\BibitemShut
  {NoStop}%
\bibitem [{\citenamefont {Volz}\ \emph {et~al.}(2013)\citenamefont {Volz},
  \citenamefont {Koelle},\ and\ \citenamefont {Bedford}}]{Volz2013}%
  \BibitemOpen
  \bibfield  {author} {\bibinfo {author} {\bibfnamefont {E.~M.}\ \bibnamefont
  {Volz}}, \bibinfo {author} {\bibfnamefont {K.}~\bibnamefont {Koelle}}, \ and\
  \bibinfo {author} {\bibfnamefont {T.}~\bibnamefont {Bedford}},\ }\bibfield
  {title} {\enquote {\bibinfo {title} {{Viral Phylodynamics}},}\ }\href
  {\doibase 10.1371/journal.pcbi.1002947} {\bibfield  {journal} {\bibinfo
  {journal} {PLOS Comput. Biol.}\ }\textbf {\bibinfo {volume} {9}},\ \bibinfo
  {pages} {e1002947} (\bibinfo {year} {2013})}\BibitemShut {NoStop}%
\bibitem [{\citenamefont {Bouckaert}\ \emph {et~al.}(2019)\citenamefont
  {Bouckaert}, \citenamefont {Vaughan}, \citenamefont {Barido-Sottani},
  \citenamefont {Duch{\^{e}}ne}, \citenamefont {Fourment}, \citenamefont
  {Gavryushkina}, \citenamefont {Heled}, \citenamefont {Jones}, \citenamefont
  {K{\"{u}}hnert}, \citenamefont {{De Maio}}, \citenamefont {Matschiner},
  \citenamefont {Mendes}, \citenamefont {M{\"{u}}ller}, \citenamefont
  {Ogilvie}, \citenamefont {du~Plessis}, \citenamefont {Popinga}, \citenamefont
  {Rambaut}, \citenamefont {Rasmussen}, \citenamefont {Siveroni}, \citenamefont
  {Suchard}, \citenamefont {Wu}, \citenamefont {Xie}, \citenamefont {Zhang},
  \citenamefont {Stadler},\ and\ \citenamefont {Drummond}}]{Bouckaert2019}%
  \BibitemOpen
  \bibfield  {author} {\bibinfo {author} {\bibfnamefont {R.}~\bibnamefont
  {Bouckaert}}, \bibinfo {author} {\bibfnamefont {T.~G.}\ \bibnamefont
  {Vaughan}}, \bibinfo {author} {\bibfnamefont {J.}~\bibnamefont
  {Barido-Sottani}}, \bibinfo {author} {\bibfnamefont {S.}~\bibnamefont
  {Duch{\^{e}}ne}}, \bibinfo {author} {\bibfnamefont {M.}~\bibnamefont
  {Fourment}}, \bibinfo {author} {\bibfnamefont {A.}~\bibnamefont
  {Gavryushkina}}, \bibinfo {author} {\bibfnamefont {J.}~\bibnamefont {Heled}},
  \bibinfo {author} {\bibfnamefont {G.}~\bibnamefont {Jones}}, \bibinfo
  {author} {\bibfnamefont {D.}~\bibnamefont {K{\"{u}}hnert}}, \bibinfo {author}
  {\bibfnamefont {N.}~\bibnamefont {{De Maio}}}, \bibinfo {author}
  {\bibfnamefont {M.}~\bibnamefont {Matschiner}}, \bibinfo {author}
  {\bibfnamefont {F.~K.}\ \bibnamefont {Mendes}}, \bibinfo {author}
  {\bibfnamefont {N.~F.}\ \bibnamefont {M{\"{u}}ller}}, \bibinfo {author}
  {\bibfnamefont {H.~A.}\ \bibnamefont {Ogilvie}}, \bibinfo {author}
  {\bibfnamefont {L.}~\bibnamefont {du~Plessis}}, \bibinfo {author}
  {\bibfnamefont {A.}~\bibnamefont {Popinga}}, \bibinfo {author} {\bibfnamefont
  {A.}~\bibnamefont {Rambaut}}, \bibinfo {author} {\bibfnamefont
  {D.}~\bibnamefont {Rasmussen}}, \bibinfo {author} {\bibfnamefont
  {I.}~\bibnamefont {Siveroni}}, \bibinfo {author} {\bibfnamefont {M.~A.}\
  \bibnamefont {Suchard}}, \bibinfo {author} {\bibfnamefont {C.-H.}\
  \bibnamefont {Wu}}, \bibinfo {author} {\bibfnamefont {D.}~\bibnamefont
  {Xie}}, \bibinfo {author} {\bibfnamefont {C.}~\bibnamefont {Zhang}}, \bibinfo
  {author} {\bibfnamefont {T.}~\bibnamefont {Stadler}}, \ and\ \bibinfo
  {author} {\bibfnamefont {A.~J.}\ \bibnamefont {Drummond}},\ }\bibfield
  {title} {\enquote {\bibinfo {title} {{BEAST 2.5: An advanced software
  platform for Bayesian evolutionary analysis}},}\ }\href {\doibase
  10.1371/journal.pcbi.1006650} {\bibfield  {journal} {\bibinfo  {journal}
  {PLOS Comput. Biol.}\ }\textbf {\bibinfo {volume} {15}},\ \bibinfo {pages}
  {e1006650} (\bibinfo {year} {2019})}\BibitemShut {NoStop}%
\bibitem [{\citenamefont {Gardy}\ \emph {et~al.}(2015)\citenamefont {Gardy},
  \citenamefont {Loman},\ and\ \citenamefont {Rambaut}}]{Gardy2015}%
  \BibitemOpen
  \bibfield  {author} {\bibinfo {author} {\bibfnamefont {J.}~\bibnamefont
  {Gardy}}, \bibinfo {author} {\bibfnamefont {N.~J.}\ \bibnamefont {Loman}}, \
  and\ \bibinfo {author} {\bibfnamefont {A.}~\bibnamefont {Rambaut}},\
  }\bibfield  {title} {\enquote {\bibinfo {title} {{Real-time digital pathogen
  surveillance — the time is now}},}\ }\href {\doibase
  10.1186/s13059-015-0726-x} {\bibfield  {journal} {\bibinfo  {journal} {Genome
  Biol.}\ }\textbf {\bibinfo {volume} {16}},\ \bibinfo {pages} {155} (\bibinfo
  {year} {2015})}\BibitemShut {NoStop}%
\bibitem [{\citenamefont {{Van Puyvelde}}\ and\ \citenamefont
  {Argimon}(2019)}]{VanPuyvelde2019}%
  \BibitemOpen
  \bibfield  {author} {\bibinfo {author} {\bibfnamefont {S.}~\bibnamefont {{Van
  Puyvelde}}}\ and\ \bibinfo {author} {\bibfnamefont {S.}~\bibnamefont
  {Argimon}},\ }\bibfield  {title} {\enquote {\bibinfo {title} {{Sequencing in
  the time of Ebola}},}\ }\href {\doibase 10.1038/s41579-018-0130-0} {\bibfield
   {journal} {\bibinfo  {journal} {Nat. Rev. Microbiol.}\ }\textbf {\bibinfo
  {volume} {17}},\ \bibinfo {pages} {5} (\bibinfo {year} {2019})}\BibitemShut
  {NoStop}%
\bibitem [{\citenamefont {Grubaugh}\ \emph {et~al.}(2019)\citenamefont
  {Grubaugh}, \citenamefont {Ladner}, \citenamefont {Lemey}, \citenamefont
  {Pybus}, \citenamefont {Rambaut}, \citenamefont {Holmes},\ and\ \citenamefont
  {Andersen}}]{Grubaugh2019}%
  \BibitemOpen
  \bibfield  {author} {\bibinfo {author} {\bibfnamefont {N.~D.}\ \bibnamefont
  {Grubaugh}}, \bibinfo {author} {\bibfnamefont {J.~T.}\ \bibnamefont
  {Ladner}}, \bibinfo {author} {\bibfnamefont {P.}~\bibnamefont {Lemey}},
  \bibinfo {author} {\bibfnamefont {O.~G.}\ \bibnamefont {Pybus}}, \bibinfo
  {author} {\bibfnamefont {A.}~\bibnamefont {Rambaut}}, \bibinfo {author}
  {\bibfnamefont {E.~C.}\ \bibnamefont {Holmes}}, \ and\ \bibinfo {author}
  {\bibfnamefont {K.~G.}\ \bibnamefont {Andersen}},\ }\bibfield  {title}
  {\enquote {\bibinfo {title} {{Tracking virus outbreaks in the twenty-first
  century}},}\ }\href {\doibase 10.1038/s41564-018-0296-2} {\bibfield
  {journal} {\bibinfo  {journal} {Nat. Microbiol.}\ }\textbf {\bibinfo {volume}
  {4}},\ \bibinfo {pages} {10--19} (\bibinfo {year} {2019})}\BibitemShut
  {NoStop}%
\bibitem [{\citenamefont {Hellewell}\ \emph {et~al.}(2020)\citenamefont
  {Hellewell}, \citenamefont {Abbott}, \citenamefont {Gimma}, \citenamefont
  {Bosse}, \citenamefont {Jarvis}, \citenamefont {Russell}, \citenamefont
  {Munday}, \citenamefont {Kucharski}, \citenamefont {Edmunds}, \citenamefont
  {{CMMID nCoV working group}}, \citenamefont {Funk},\ and\ \citenamefont
  {Eggo}}]{Hellewell2020}%
  \BibitemOpen
  \bibfield  {author} {\bibinfo {author} {\bibfnamefont {J}~\bibnamefont
  {Hellewell}}, \bibinfo {author} {\bibfnamefont {S}~\bibnamefont {Abbott}},
  \bibinfo {author} {\bibfnamefont {A}~\bibnamefont {Gimma}}, \bibinfo {author}
  {\bibfnamefont {NI}~\bibnamefont {Bosse}}, \bibinfo {author} {\bibfnamefont
  {CI}~\bibnamefont {Jarvis}}, \bibinfo {author} {\bibfnamefont
  {TW}~\bibnamefont {Russell}}, \bibinfo {author} {\bibfnamefont
  {JD}~\bibnamefont {Munday}}, \bibinfo {author} {\bibfnamefont
  {AJ}~\bibnamefont {Kucharski}}, \bibinfo {author} {\bibfnamefont
  {WJ}~\bibnamefont {Edmunds}}, \bibinfo {author} {\bibnamefont {{CMMID nCoV
  working group}}}, \bibinfo {author} {\bibfnamefont {S}~\bibnamefont {Funk}},
  \ and\ \bibinfo {author} {\bibfnamefont {RM}~\bibnamefont {Eggo}},\
  }\bibfield  {title} {\enquote {\bibinfo {title} {Feasibility of controlling
  2019-ncov outbreaks by isolation of cases and contacts},}\ }\href
  {https://cmmid.github.io/ncov/isolation_contact_tracing} {\bibfield
  {journal} {\bibinfo  {journal} {{online preprint}}\ } (\bibinfo {year}
  {2020})}\BibitemShut {NoStop}%
\bibitem [{\citenamefont {Newman}(2012)}]{Newman2012b}%
  \BibitemOpen
  \bibfield  {author} {\bibinfo {author} {\bibfnamefont {M.~E.~J.}\
  \bibnamefont {Newman}},\ }\href@noop {} {\emph {\bibinfo {title}
  {{Computational Physics}}}}\ (\bibinfo  {publisher} {CreateSpace Independent
  Publishing Platform},\ \bibinfo {year} {2012})\ p.\ \bibinfo {pages}
  {562}\BibitemShut {NoStop}%
\end{thebibliography}
\end{document}